\documentclass[referee]{raa}
\usepackage{txfonts}
\usepackage{graphicx,times}
\usepackage{natbib}
\usepackage{amssymb}
\usepackage{threeparttable}
\usepackage{longtable}

\newcommand{\Msun}[0]{M_{\sun}}
\newcommand{\Rsun}[0]{R_{\sun}}
\newcommand{\Lsun}[0]{L_{\sun}}
\newcommand{\Gyr}[0]{\rm{Gyr}}
\newcommand{\dex}[0]{\rm{dex}}

\begin{document}

   \title{Stellar Parameters of Main Sequence Turn-off Star Candidates Observed with the LAMOST and \emph{Kepler}}

   \volnopage{Vol.0 (200x) No.0, 000--000}      
   \setcounter{page}{1}          

   \author{Yaqian Wu
      \inst{1}
      \and Maosheng Xiang
      \inst{2}
      \and Xianfei Zhang
      \inst{1}
      \and Tanda Li
      \inst{2}
      \and Shaolan Bi
      \inst{1}
      \and Xiaowei Liu
      \inst{3}
      \and Jianning Fu
      \inst{1}
      \and Yang Huang
      \inst{3}
      \and Zhijia Tian
      \inst{3}
      \and Kang Liu
      \inst{1}
   \and Zhishuai Ge
      \inst{1}
   \and Xin He
      \inst{1}
  \and Jinghua Zhang
      \inst{1}
   }

   \institute{Department of Astronomy, Beijing Normal University,
             Beijing 100875, P.\ R.\ China; {\it wuyaqian@mail.bnu.edu.cn; bisl@bnu.edu.cn}\\
             \and
             National Astronomical Observatories, Chinese Academy of Sciences,
             Beijing 100012, P.\ R.\ China\\
             \and
             Department of Astronomy, Peking University,
             Beijing 100871, P.\ R.\ China\\
             }

   \date{Received~~2015 month day; accepted~~2015~~month day}

\abstract{Main sequence turn-off (MSTO) stars have advantages as indicators of Galactic evolution since their ages could be robustly estimated from atmospheric parameters. Hundreds of thousands of MSTO stars have been selected from the LAMOST Galactic survey to study the evolution of the Galaxy, and it is vital to derive accurate stellar parameters. In this work, we select 150 MSTO star candidates from the MSTO stars sample of Xiang that have asteroseismic parameters and determine accurate stellar parameters for these stars combing the asteroseismic parameters deduced from the $Kepler$ photometry and atmospheric parameters deduced from the LAMOST spectra. With this sample, we examine the age determination as well as the contamination rate of the MSTO stars sample. A comparison of age between this work and Xiang shows a mean difference of 0.53\,Gyr (7\%) and a dispersion of 2.71\,Gyr (28\%). The results show that 79 of the candidates are MSTO stars, while the others are contaminations from either main sequence or sub-giant stars. The contamination rate for the oldest stars is much higher than that for the younger stars. The main cause for the high contamination rate is found to be the relatively large systematic bias in the LAMOST surface gravity estimates.
}

   \authorrunning{Stellar Parameters of MSTO Star Candidates Observed by LAMOST and \emph{Kepler}}
   \titlerunning{Y.Q.Wu et al.}

   \maketitle

\section{INTRODUCTION}
\label{sect:intro}
A star begins its evolution as a hydrogen-rich main-sequence star with a hydrogen-burning core.
As core hydrogen burning finishes, hydrogen-shell burning starts and the star expands to larger radius, lower surface temperature and higher luminosity,
and the star evolves into the sub-giant branch phase.
Main sequence turn-off (MSTO) stars are stars that have reached the point of central hydrogen exhaustion at the end of the main-sequence phase. Given the metallicity, their effective temperatures are very sensitive to their ages, hence one can obtain reliable age estimates for MSTO stars given accurate measurements of effective temperatures.

 MSTO stars are widely used to determine ages of star clusters as they are easily identified from the color-magnitude diagrams (CMDs) \citep[e.g.][]{Mackey08, Goudfrooij09, Yang13}. Since member stars of a cluster are generally believed to form from the same gas cloud simultaneously, they have the same age.
Unlike the MSTO stars in clusters, MSTO stars in the field are not easy to be identified
from the CMDs. To identify field MSTO stars, accurate estimates of atmospheric parameters ($T_{\rm eff}$,
 $\log g$ and [Fe/H] ) are required. As the implementation of the LAMOST Experiment for Galactic Understanding and Exploration \citep[LEGUE;][]{Deng12, Zhao12, Liu14} and other spectroscopic surveys such as the Sloan Extension for Galactic Understanding and Exploration \citep[SDSS/SEGUE;][]{Yanny09}, the Radial Velocity Experiment \citep[RAVE; ][]{Steinmetz06} and the Apache Point Observatory Galactic Evolution Experiment \citep[APOGEE;][]{Majewski10}, stellar atmospheric parameters of millions of stars are delivered from the survey spectra. Typical accuracies of the LAMOST stellar atmospheric parameters reach 100 -- 150\,K for  $T_{\rm eff}$, 0.20 -- 0.25\,dex for log\,$g$ and 0.1 -- 0.2\,dex for [Fe/H] (Xiang et al. 2015b, Wu et al. 2014, Luo et al. 2015, Gao et al. 2015). Hundreds of thousands of MSTO stars have been selected from the LAMOST survey by Xiang et al. (2015a) 
 based on stellar atmospheric parameters yielded by the LAMOST Stellar Parameter Pipeline at Peking University (LSP3; Xiang et al. 2015b). The ages of these MSTO stars are also estimated, with a claimed uncertainty of about 30 per cent.
However, given the low spectral resolving power of LAMOST ($R\sim$1800; e.g. Cui et al. 2012; Deng et al. 2012), accurate stellar parameters, especially surface gravity, are difficult to yield from the spectra. Therefore, a careful sanity examination on the feasibility of the method to select the MSTO stars sample and on the accuracy of age estimation seems to be essential.

Asteroseismology is a powerful tool to derive accurate stellar parameters \citep{Bi08,G10, Yang10, Chaplin11, Stello13, Tian15}. By asteroseismology, accurate stellar parameters of thousands of stars have been obtained \citep{Chaplin14, Huber14}.
It is found that surface gravities yielded by the asteroseismology can be accurate to 0.01 -- 0.03\,dex \citep{Hekker13, Huber14}, much better than the spectroscopic estimates. Combing effective temperatures and metallicities from the LAMOST spectra with asteroseismic surface gravity yielded from the $Kepler$ photometry, MSTO stars can be well identified and their ages can also be accurately determined.

In this paper, we determine fundamental stellar parameters ($M$, $R$, Age, $L$, $T_{\rm eff}$, $Z$, $\log g$) for 150 MSTO star candidates selected from the MSTO stars sample that have asteroseismic properties delivered from photometry of the $Kepler$ mission \citep{G10}. Meanwhile, we compare our results with previous studies by Huber et al.\ (2014) and Xiang et al.\ (2015a). We discuss the impact of uncertainties in atmospheric parameters on the measurement of ages, and examine the accuracy of the age estimates as well as the contamination rate of the MSTO stars sample. The paper is organized as follows. In Section\,2, we introduce how to select the sample of MSTO star candidates. In Section\,3, we describe the stellar model and how to obtain the stellar parameters. Results and discussion are presented in Section\,4 and a summary is shown in Section\,5.

\section{THE MSTO STAR CANDIDATES}
\label{sect:Dat}
The LAMOST-\emph{Kepler} project \citep{De Cat Peter15} aims to observe
stars in the \emph{Kepler} field with the LAMOST \citep{Cui12} and deliver atmospheric parameters and radial velocities.
The LAMOST survey \citep{Zhao12} has produced a large number of low
resolution ($R$$\sim$1800) optical spectra ($\lambda$\,3800 -- 9000\,${\AA}$).
By September 2014, all the \emph{Kepler} fields had been observed at least once,
and $101\,086$ spectra had beeen collected on 38 LAMOST plates \citep{De Cat Peter15}.
Many of the stars have asteroseismic characteristics deduced from the \emph{Kepler} photometry.
These stars have been used to examine and calibrate stellar surface gravities
yielded from the LAMOST spectra \citep{Ren16, Wang16}.

To study the evolution of the Galaxy, Xiang et al. (2015a) have selected a sample of 0.3 million MSTO stars from the LAMOST survey based on stellar atmospheric parameters ($T_{\rm eff}$, log\,$g$ and [Fe/H]) derived by the LSP3. They have estimated stellar ages for those MSTO stars with a isochrone fitting technique utilizing the Yonsei-Yale (YY) isochrones (Demarque et al. 2004), and claim a typical accuracy of 30 per cent.

Among the LAMOST-$Kepler$ stars, about 4000 stars are found to have asteroseismic parameters the frequency of maximum power of the oscillation $\nu_{\rm max}$ and large frequency spacing $\Delta\nu$) from literature \citep{Hekker11, Appourchaux, Mosser12, Stello13, Huber13, Chaplin14}, most of which are red giant stars and only about 300 stars are dwarfs/subgiants. A cross-identification with the MSTO stars sample yields 179 common stars, and in this work, we denote them as MSTO star candidates. Asteroseismic parameters collected from literature, as well as atmospheric parameters yielded by the LSP3 for these MSTO star candidates are listed in Table\,1.

\section{GRID MODELING}
\label{sect:Mod}
\subsection{Models}
\label{3.1}
We use the Yale Rotating Stellar Evolution Code \citep[YREC;][]{Pin90, Pin92, Demarque08} to construct stellar evolution models.
Input physics include the OPAL equation of state tables \citep{Rogers02}
and OPAL high-temperature opacities \citep{Iglesias96} supplemented with low-temperature opacities of Ferguson et al.\ (2005).
The NACRE nuclear reaction rates \citep{Bahacall95} are used.
Atomic diffusion due to concentration and thermal gradients is included in
the computation of models with initial masses below $1.1\,\Msun$, using the formulation of Thoul et al.\ \cite{Thoul94}.
For grids with initial mass between 1.1 and $1.2\,\Msun$, both models with and without atomic diffusion are calculated.
The outer convective zone is treated according to the mixing-length theory \citep{Bohm58}
and the influence of overshooting convection is ignored.
To account for the uncertain mixing-length parameter, $\alpha_{\rm MLT}$,
three sets of the model grids are calculated, each with an $\alpha_{\rm MLT}$ of 1.75, 1.84 and 1.95,
respectively; the solar-calibrated value is $\alpha_{\rm MLT}=1.84$.

We calculate stellar evolution models with [Fe/H] in the range $-0.3$ -- $0.4\,\dex$ in steps of $0.1\,\dex$.
We assume that $[\rm{Fe/H}] = 0$ corresponds to the solar abundance ($(Z/X)_{\sun}=0.0231$)
as determined by Grevesse et al.\ \cite{Grevesse98} and that these models have a helium abundance of $Y = 0.248$.
The helium abundance for models with other values of metallicity
is determined assuming a chemical evolution model $Y_{\rm ini}=0.248 + \Delta Y/\Delta Z \times Z$, where $\Delta Y/\Delta Z$=1.33.
For each [Fe/H], the ratio of heavy-elements to hydrogen as a mass fraction is estimated through the formula
\begin{equation}
[\rm{Fe/H}]= \log  \left( \frac{Z}{X} \right) - \log \left( \frac{Z}{X} \right)_{\sun}.
\end{equation}
Our models have masses in the range $0.8$--$2.5\,\Msun$, in steps of $0.02\,\Msun$.
The evolution tracks are constructed from the pre-main sequence to the base of the red-giant branch (RGB).
A summary of the adopted input parameters is given in Table\,\ref{tbl3}.

Figure\,1 shows the evolutionary tracks in the $T_{\rm eff}$ -- $\log g$ plane
and the position of the $179$ MSTO star candidates, with parameters derived with the LSP3. The error bars represent the error of $T_{\rm eff}$ and $\log g$ derived from the LSP3 separately.
The figure indicates that most of the MSTO star candidates are located around the main sequence turn-off stage.

\subsection{Methodology}
Usually, stellar parameters of field MSTO stars are determined by comparing theoretical
models with atmospheric parameters such as $T_{\rm eff}$ and [Fe/H] derived from either photometry or spectroscopy.
Low-mass main-sequence stars and some sub-giants show rich spectra
of solar-like oscillations, small amplitude pulsations which are excited
and damped intrinsically by convection in the outer envelope.
The large frequency spacing, $\Delta\nu$, is formally related to the mean density of a star \citep{Christensen-Dalsgaard93}.
the frequency of maximum power of the oscillation $\nu_{\rm max}$ is related
to the acoustic cutoff frequency of a star \citep[e.g.][]{Kjeldsen95, Bedding03, Chaplin08}.
Both $\Delta\nu$ and $\nu_{\rm max}$ are sensitive to the structure of stars,
and thus are indicators of evolutionary stage.

Under the constraints of atmospheric parameters ($T_{\rm eff}$, [Fe/H]) and seismic properties ($\Delta\nu$ and $\nu_{\rm{max}}$) listed in Table~\ref{table:1}, we use the likelihood method of Basu et al. (2010) to find the best-fit models.

Given the observed and model parameters, the likelihood is:
\begin{equation}
\label{a3}
  L_{\rm{T_{\rm{eff}}}} =\frac{1}{\sqrt{2\pi} \sigma_{\rm{T_{\rm{eff}}}}}\exp\left(\frac{-(T_{\rm{eff}_{\rm{obs}}}-T_{\rm{eff}_{\rm{model}}})^2}{2\sigma_{\rm{T_{\rm{eff}}}}^2}\right),
\end{equation}

\begin{equation}
\label{a4}
  L_{\rm{[Fe/H]}} =\frac{1}{\sqrt{2\pi}\sigma_{\rm{[Fe/H]}}}\exp\left(\frac{-(\rm [Fe/H]_{\rm{obs}}-\rm [Fe/H]_{\rm{model}})^2}{2\sigma_{\rm{[Fe/H]}}^2}\right),
\end{equation}

\begin{equation}
\label{a1}
  L_{\rm{\Delta\nu}} =\frac{1}{\sqrt 2\pi\sigma_{\rm{\Delta\nu}}}\exp\left(\frac{-(\Delta\nu_{\rm{obs}}-\Delta\nu_{\rm{model}})^2}{2\sigma_{\rm{\Delta\nu}}^{2}}\right),
\end{equation}

\begin{equation}
\label{a2}
  L_{\rm\nu_{\rm{max}}} =\frac{1}{\sqrt 2\pi\sigma_{\rm\nu_{\rm{max}}}}\exp\left(\frac{-(\nu_{\rm{max,obs}}-\nu_{\rm{max,model}})^2}{2\sigma_{\rm{\nu_{\rm{max}}}}^2}\right).
\end{equation}
 The combined likelihood is
 \begin{equation}
\label{a5}
L=L_{\rm{\Delta\nu}}\ L_{\rm\nu_{\rm{max}}}\ L_{\rm{T_{\rm{eff}}}}\ L_{\rm{[Fe/H]}}.
\end{equation}
 Note that we do not consider the likelihood function for $\log g$
 because the LSP3 estimates of this quantity may have large systematic errors.
We assume that the normalized probability of each model $p_{i}$ is:
 \begin{equation}
p_{i}=\frac{L_{i}}{\sum_{i=1}^{N_{\rm m}}L_{i}},
\end{equation}
where $N_{\rm m}$ is the total number of models.
This normalized probability is a measurement of how well the
each model in the set matches the parameters of an observed star. Similar to Kallinger et al. (2010), we use the integral probability to estimate the best-fitted parameter and its error.
For each parameter, the best-fitted parameter respects to value that have a integral probability of 0.5, and the $1 \sigma$ error is given.
\section{RESULTS AND DISCUSSION}

Among the 179 MSTO star candidates, stellar parameters for 150 of them are successfully derived, and are listed in Table\,3, while the remaining 29 stars are falling outside of our model grids. For the 29 remaining stars, 25 stars are RGB stars according to their asteroseismic characteristics
and the other 4 stars are metal-poor stars with $[\rm{Fe/H}] < -0.3\,\dex$.

Figure~2 illustrates distributions of the derived mass, age and metallicity for the 150 MSTO star candidates.
Their masses are in the range of $0.8$ -- $1.5\,\Msun$
and peak at about $1.1\,\Msun$.
The ages are widely distributed in the range of $0$ -- $13\,\Gyr$,
mostly in $2$ -- $8\,\Gyr$.
The metallicities distributed in the range of $-0.3$ -- $0.3\,\dex$,
with a moderate peak near the solar value. Besides,
the typical uncertainties in $T_{\rm eff}$, $\log g$, [Fe/H], $M$
and $R$ are $60\,\rm{K}$, $0.009\,\dex$, $0.1\,\dex$, $0.04\,\Msun$ and $0.03\,\Rsun$, respectively.
The uncertainty in $\log g$ is much smaller than that yielded by the LSP3 from the LAMOST spectra.
And uncertainties in the stellar age vary from $0.4\,\Gyr$ for
young stars to $1.3\,\Gyr$ for old stars, corresponding to a relative error about $9$ -- $10\,\%$.

In fact, part of the stellar parameters for those $150$ MSTO star candidates are also provided by Huber et al. ~(2014). Based on the Dartmouth Stellar Evolution Database \citep[DSEP,][]{Dotter08}, Huber et al. (2014) derived fundamental parameters using the asteroseismic quantities and atmospheric parameters in the $Kepler$ input catalog (KIC), which are estimated from photometry for most stars. But they did not provide ages. 
By comparing the $\log g$, $M$ and $R$ derived by Huber et al.\ (2014) and those of our work, we find that our estimates of $\log g$ are consistent with those of Huber et al. very well,
which yield a mean difference of only $0.0002\,\dex$,
and a standard deviation of $0.013\,\dex$.
However, there are systematic deviations of $R$ and $M$ between our estimates and those
of Huber et al. Our values are systematically smaller than those of Huber et al., and the deviations increase with increasing $R$ and $M$. Typical difference of $R$ is 0.025 -- 0.05\,$R_{\odot}$ for stars with $R$\ $\sim$\ 2.0  $R_{\odot}$, and typical difference of $M$ is 0.1 -- 0.2\,$M_{\odot}$ for stars with $M$\,$\sim$\,1.5 $M_{\odot}$. Nevertheless, the dispersion of differences of $R$ and $M$ are small (after exclude systematic trends), with only 0.056\,$R_{\odot}$ in $R$, and 0.1\,$M_{\odot}$ in $M$.
The systematic trends of differences in $R$ and $M$ are mainly caused by systematic differences in the effective temperatures adopted.
In Figure\,4, we compare $T_{\rm eff}$ derived by the LSP3 and those of Huber et al.
The figure exhibits similar trends to those for the mass in Figure\,3.
As expected, there is strong correlation between the differences of mass and the differences of effective temperatures.
Because the $T_{\rm eff}$ derived by the LSP3 are calibrated to the recently
deduced metallicity-dependent color--temperature relation of Huang et al. \citep{15},
which is deduced based on stellar interferometry data sets,
we believe our results of stellar mass are more reliable than those of Huber et al.
In addition, the stellar metallicity could also affect the determination of stellar mass,
but it has only a minor contribution compared to that from the effective temperature.

After that, in Figure\,5, we compare age estimates of this work with those of Xiang et al. (2015a). 
The left panel shows that though ages for the majority of stars agree well with each other, there are a considerable fraction of stars that our values are systematically much lower than those of Xiang et al.  For stars older than 10\,Gyr based on Xiang et al.'s age estimates, about half of them are actually younger than 7\,Gyr according to our results. 
The right panel plots the distribution of the differences of age estimates.
The distribution yields a mean difference of $0.53\,\Gyr$ ($7\,\%$),
and a dispersion of $2.71\,\Gyr$ ($28\,\%$).
It is found that the age discrepancy are mostly caused by systematic bias in the LSP3 $\log g$.
For instance, KIC 5523099, the LSP3 atmospheric parameters ($5513\,\rm{K}$, $4.24\,\dex$, $0.03\,\dex$)
yield an age of $12.5\,\Gyr$,
while our atmospheric parameters ($5507\,\rm{K}$, $3.79\,\dex$, $0.05\,\dex$)
yield $4.6\,\Gyr$, which is $6.9\,\Gyr$ younger due to a $0.45\,\dex$ overestimate of the LSP3 $\log g$.
As the uncertainty of the LSP3 $\log g$ is the main cause of the differences in stellar ages, left panel in Figure\,6, we compare $\log g$ derived by the LSP3 with our values. The figure reveals that $\log g$ given by the LSP3 has a linear trend of deviation with our estimated values.
The result is consistent with that of Ren et al.~(2016), who examined the LSP3 $\log g$ with asteroseismic values from Huber et al. (2014).
To better characterize the bias in the LSP3 $\log g$, we display the histogram distribution of log\ $g$ differences in right panel in Figure\,6. The figure exhibits that the LSP3 $\log g$ is generally higher
than our seismic values by about $0.1\,\dex$, with a calculated standard deviation of $0.16\,\dex$.

We compare the $T_{\rm eff}$ -- $\log g$ diagram of the LSP3 and our work in Figure\,7 .
The figure indicates that our work yields a sparser distribution, and that
a considerable fraction of the stars are located in the sub-giant branch.
Based on the definition of the MSTO stars of Xiang et al.(2015a), our revised atmospheric parameters indicate that only $79$ of the $150$ star candidates are MSTO stars, while the other $71$ are contaminations from either main sequence or sub-giant stars. The stellar ages for those 71 contamination stars are marked with red circles in Figure\,5. Considering the $29$ stars falling outside of our model grids, which 4 metal-poor ones maybe MSTO stars and the other 25 stars are certainly RGB stars, there are $46\,\%$ (83/179) stars in total are MSTO stars, while the others are contaminations from either main sequence or sub-giant stars. However, considering that the number of stars in our sample is still small, and that the asteroseismic sample from literature are probably biased to sub-giant stars because they are brighter and also have relatively larger oscillation amplitudes thus easier to be detected, our results have probably overestimated the contamination rate.

\section{SUMMARY}
 Combing atmospheric parameters derived with the LSP3 from the LAMOST spectra, and seismic characteristics derived from \emph{Kepler} photometry, we have determined the stellar parameters for 150 MSTO star candidates selected from the MSTO stars sample by constructing stellar evolution models.
Typical uncertainties for their parameters are $0.04\,\Msun$, $0.03\,\Lsun$, $0.03\,\Rsun$ for $M$, $L$ and $R$, respectively,
$0.4\,\Gyr$ for young stars and $1.3\,\Gyr$ for old stars,
as well as $60\,\rm{K}$, $0.009\,\dex$, $0.1\,\dex$ for $T_{\rm eff}$, $\log g$ and [Fe/H], respectively.

Meanwhile,we compare the derived $\log g$, radius and mass with those of Huber et al.\ (2014), and
find that the $\log g$ and radius are consistent well with each other, while the mass show moderate differences due to different effective temperatures adopted.
We also compare our ages estimates with those of Xiang et al.\ (2015a) and find a mean difference of $0.53\,\Gyr$ ($7\,\%$) and a dispersion of $2.71\,\Gyr$ ($28\,\%$). Moreover, we also re-select MSTO stars based on the criteria of Xiang et al. (2015a) utilizing our newly derived atmospheric parameters and find that about half of the MSTO stars identified with the LSP3 atmospheric parameters are actually main sequence or sub-giant stars, and the stellar ages for those contamination stars are systematically overestimated. The contamination is especially dramatic for oldest stars in the MSTO stars sample. However, the number of stars in our sample is still small, , and they are probably biased to sub-giant stars, so that our sample may not be representative enough to give a full clarification of contamination rate of the MSTO stars sample. As the LAMOST survey progresses, we plan to obtain larger sample to deduce more conclusive results in the next.

\begin{acknowledgements}
Guoshoujing Telescope (the Large Sky Area Multi-Object Fiber Spectroscopic Telescope LAMOST)
is a National Major Scientific Project built by the Chinese Academy of Sciences.
Funding for the project has been provided by the National Development and Reform Commission.
LAMOST is operated and managed by the National Astronomical Observatories, Chinese Academy of Sciences.
This work is supported by grants 11273007 and 10933002 from the National Natural Science Foundation of China, the Joint Research Fund in Astronomy (U1631236) under cooperative agreement between the National Natural Science Foundation of China (NSFC) and Chinese Academy of Sciences (CAS), and the Fundamental Research Funds for the Central Universities and Youth Scholars Program of Beijing Normal University.
\end{acknowledgements}

\clearpage

\begin{figure}
\centering
\includegraphics[width=120mm, angle=0]{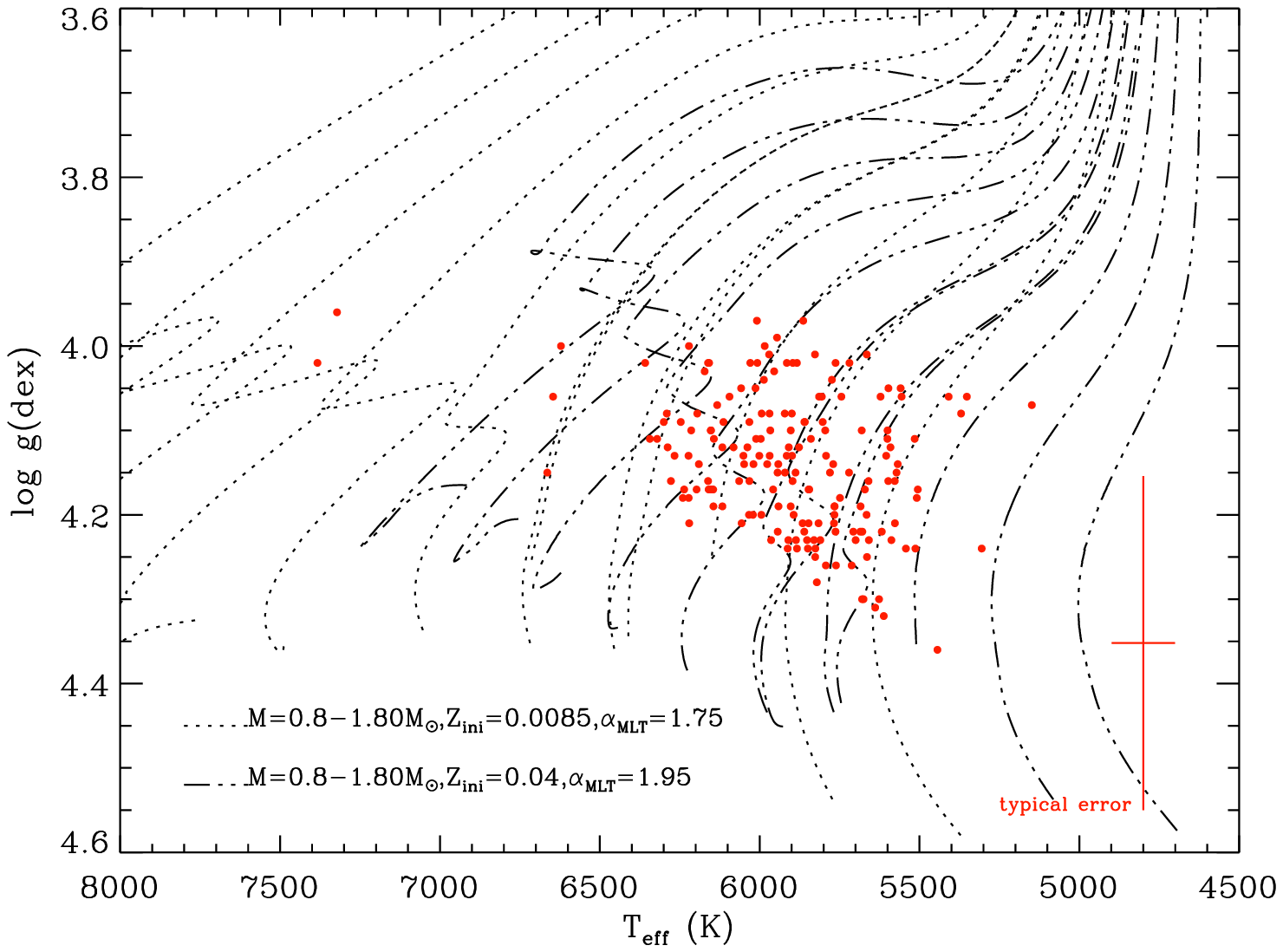}
\caption{MSTO stars and evolution tracks in the $T_{\rm eff}$--$\log g$ plane.
The black dotted lines and dashed lines represent the stellar tracks with masses
between $0.8$ and $1.8\,\Msun$, $Z_{\rm ini}$ between $0.0085$ and $0.04$ and $\alpha_{\rm MLT}$ between $1.75$ and $1.95$ respectively.
The red filled circles represent the 179 MSTO stars with parameters derived by the LSP3, listed in Table~1, typical error bars are presented in the bottom-right corner of the figure.}
\label{Fig:1}
\end{figure}

\begin{figure}
\centering
\includegraphics[width=100mm, angle=0]{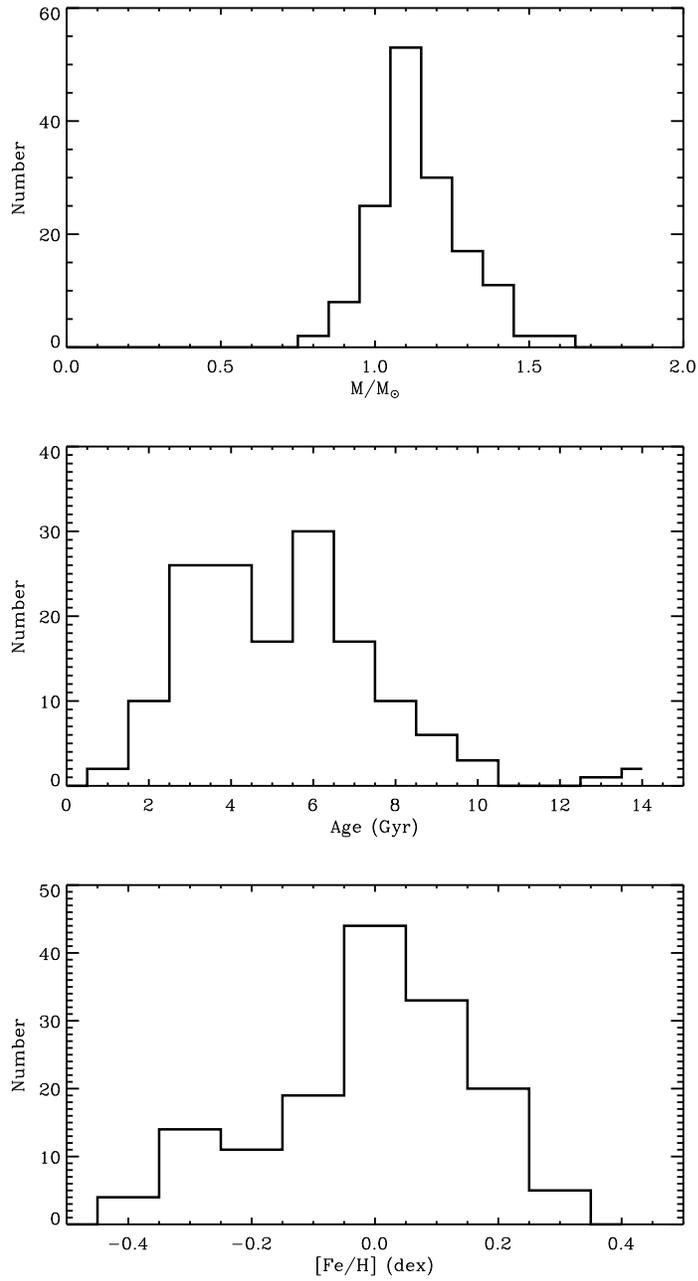}
\caption{Distribution of 150 MSTO star with revised parameters.
Different parameters lies in different panels, i.e.,
top: the distribution of masses,
middle: the distribution of ages, bottom: distribution of metallicities.}
\label{Fig:2}
\end{figure}

\begin{figure}
\centering
\includegraphics[width=180mm, angle=0]{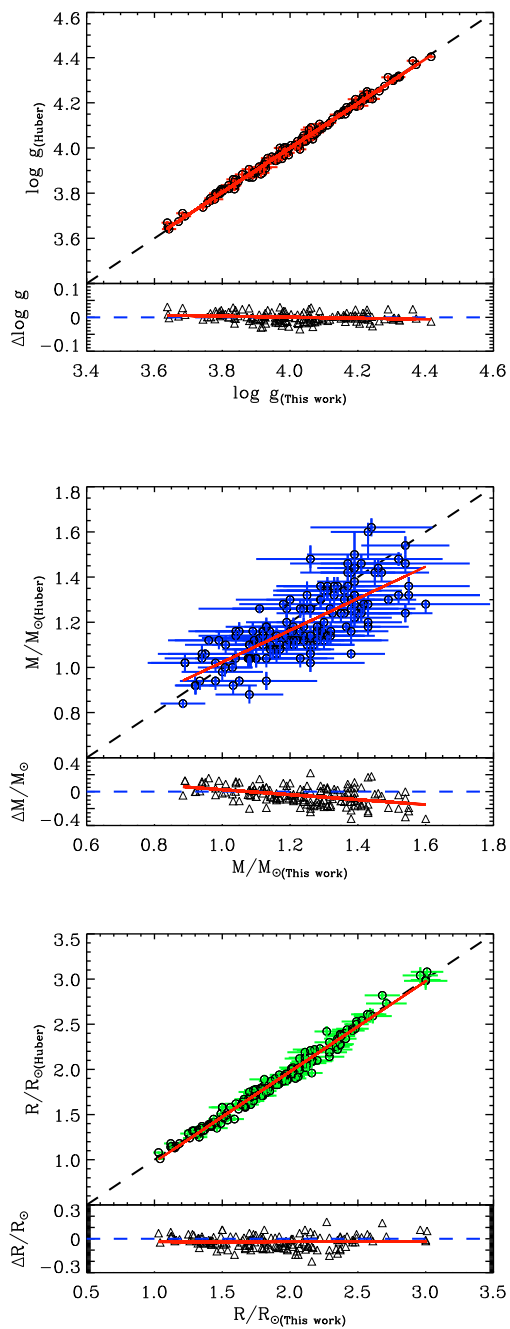}
\caption{Comparison of the results of Huber et al. (2014) and our work.
Different parameters lies in different panels, i.e.,
top: gravity $\log g$, middle: mass $M$, bottom: radius $R$.
The dashed line shows line of equality,
and the solid line shows a least square method fitting of both results.  }
\label{Fig:3}
\end{figure}

\begin{figure}
\centering
\includegraphics[width=100mm, angle=0]{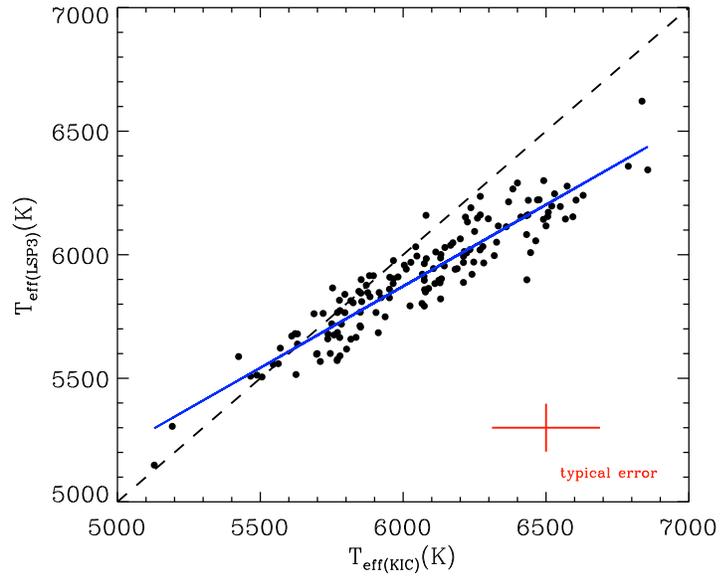}
\caption{Comparison the $T_{\rm eff}$ from LAMOST and \emph{Kepler}.
The dashed line shows line of equality.
Blue solid line shows least squares fitting of temperature. Typical error bars are presented in the bottom-right corner of the figure.}
\label{Fig:4}
\end{figure}

\begin{figure}
\centering
\includegraphics[width=160mm, angle=0]{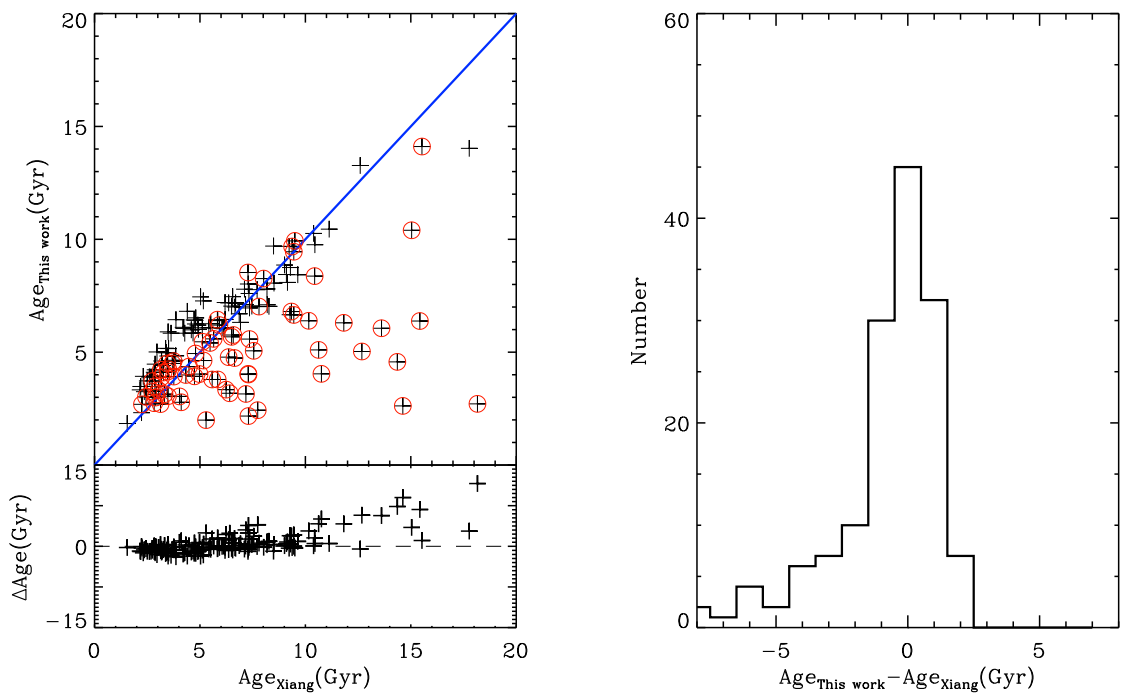}
\caption{Comparison the ages from Xiang et al.~(2015a). and our work.
Left panel: comparison of age calculated by isochrone fitting (Xiang) and asteroseismology (our work).
The blue solid line shows line of equality.
The crosses shows the sample of 150 stars and red circles indicate the non-MSTO stars in our work.
Right panel: histogram of differences of ages.}
\label{Fig:5}
\end{figure}

\begin{figure}
\centering
\includegraphics[width=180mm, angle=0]{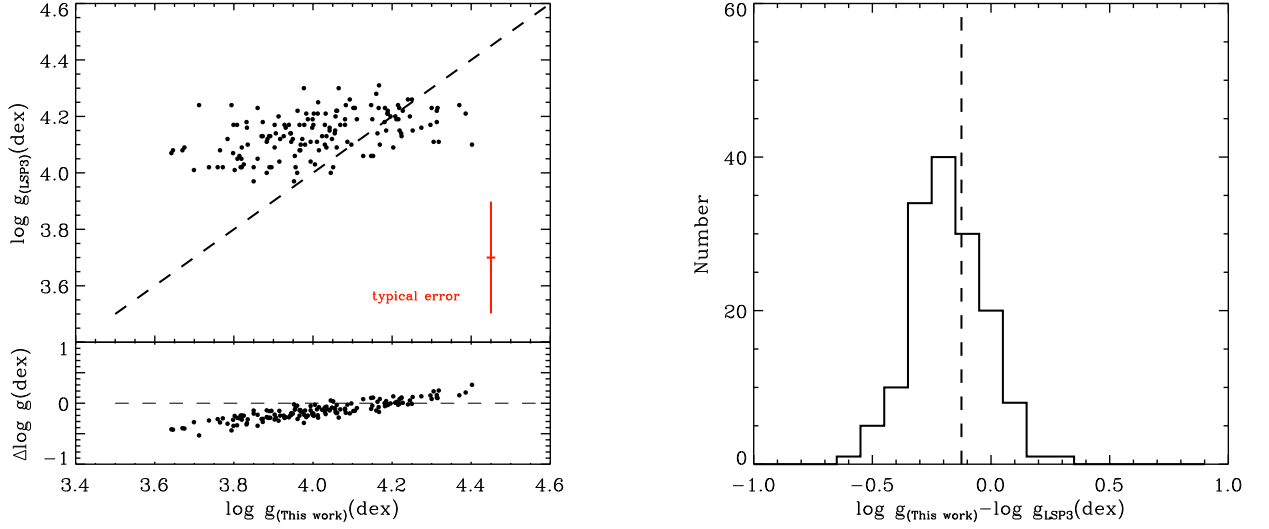}
\caption{Comparison of the $\log g$, which derived from LSP3, with our work.
Left panel: comparison of log\ $g$ between LSP3 and our work, the dashed line shows line of equality.
Right panel: distribution of the difference in $\log g$ between LSP3 (LSP3) and this work.
The dashed line shows the median value of the distribution.}
\label{Fig:6}
\end{figure}

\begin{figure}
\centering
\includegraphics[width=120mm, angle=0]{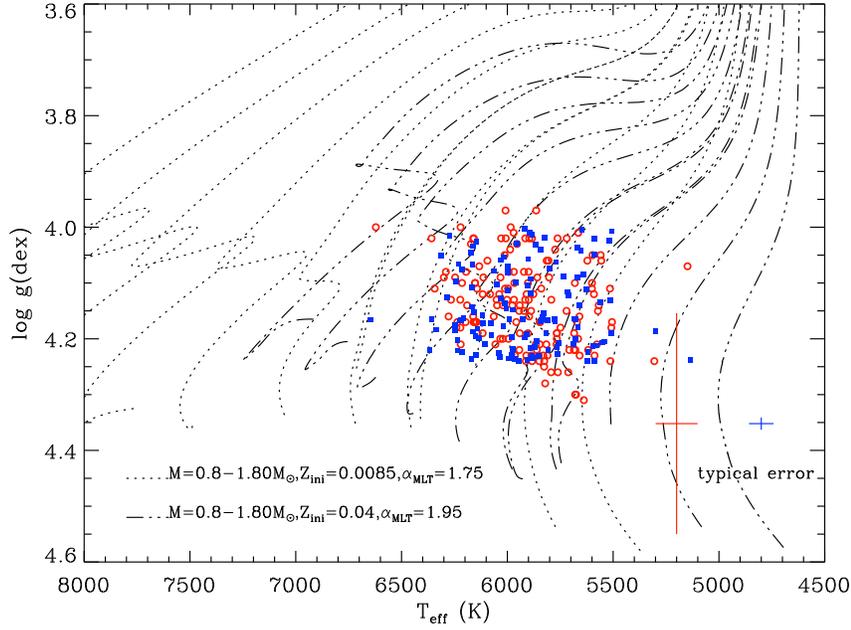}
\caption{The location of 150 MSTO candidates with parameters from this work (blue filled circles) or the LSP3 (red open circles). Typical error bars are presented in the bottom corner of the figure.
 Black dotted lines and dashed lines are evolution tracks for $M=0.8$--$1.5\,\Msun$ and $Z=0.0085$--$0.04$. }
\label{Fig:8}
\end{figure}

\clearpage

\begin{center}
\begin{longtable}{ccccccc}
    \caption{Spectroscopic and Asteroseismic parameters of 179 targets.}\label{table:1}\\
    \endfirsthead																																
\multicolumn{7}{c}%
{{\tablename\ \thetable{} -- continued from previous page}} \\ \hline																																						
 Star & ${T_{\rm eff}}$ & $\log g$ & $[\rm{Fe/H}]$  & $\Delta\nu$ & $\nu_{\rm max}$ & Ref. \\

 KIC & (K) & (dex) & (dex)  & ($\mu\rm{Hz}$) & ($\mu\rm{Hz}$) & \\
\hline																																					
\hline																																						
\endhead																																						
\hline \multicolumn{7}{c}{{Continued on next page}} \\																																						
\endfoot																																						
\hline																																						
\endlastfoot																																						
\hline																																						
 Star & ${T_{\rm eff}}$ & $\log g$ & $[\rm{Fe/H}]$  & $\Delta\nu$ & $\nu_{\rm max}$ & Ref.\\

 KIC & (K) & (dex) & (dex)  & ($\mu\rm{Hz}$) & ($\mu\rm{Hz}$) & \\
				\hline																																		
				\hline	

    1725815 & $6195\pm151$ & $4.08\pm0.23$ & $0.04\pm0.11$ & $55.4\pm1.3$ & $1045\pm47$ & (1)\\
    2010607 & $6113\pm100$ & $4.09\pm0.21$ & $0.16\pm0.09$ & $42.5\pm1.7$ & $675\pm86$ & (1)\\

    2865774 & $5792\pm93$ & $4.13\pm0.20$ & $0.07\pm0.09$ & $62.7\pm2.7$ & $1252\pm90$ & (1)\\
    2998253 & $6116\pm100$ & $4.19\pm0.19$ & $0.16\pm0.09$ & $89.0\pm3.6$ & $   $ & (1)\\
    3112152 & $5967\pm97$ & $4.10\pm0.20$ & $-0.03\pm0.09$ & $63.9\pm1.9$ & $1219\pm67$ & (1)\\
    3123191 & $6145\pm101$ & $4.19\pm0.19$ & $-0.05\pm0.10$ & $89.8\pm2.2$ & $  $ & (1)\\
    3241581 & $5577\pm89$ & $4.21\pm0.18$ & $0.27\pm0.09$ & $122.9\pm1.6$ & $  $ & (1)\\
    3344897 & $6240\pm104$ & $4.18\pm0.20$ & $0.07\pm0.10$ & $46.5\pm0.8$ & $874\pm56$ & (1)\\
    3438633 & $6082\pm100$ & $4.12\pm0.20$ & $-0.17\pm0.10$ & $55.8\pm1.7$ & $987\pm56$ & (1)\\
    3456181 & $6236\pm104$ & $4.17\pm0.20$ & $0.07\pm0.10$ & $52.0\pm0.8$ & $921\pm30$ & (1)\\

    3656476 & $5568\pm89$ & $4.14\pm0.19$ & $0.30\pm0.09$ & $93.3\pm1.3$ & $1887\pm40$ & (1)\\
    3942719 & $5659\pm90$ & $4.16\pm0.19$ & $-0.31\pm0.11$ & $45.9\pm3.7$ & $780\pm38$ & (1)\\

    3967859 & $5888\pm95$ & $4.15\pm0.20$ & $-0.27\pm0.10$ & $93.6\pm3.0$ & $   $ & (1)\\
    4049576 & $5803\pm93$ & $4.09\pm0.20$ & $-0.19\pm0.10$ & $51.9\pm1.7$ & $942\pm70$ & (1)\\
    4141376 & $5903\pm95$ & $4.10\pm0.20$ & $-0.28\pm0.11$ & $128.8\pm1.3$ & $2928\pm97$ & (2)\\
    4143755 & $5681\pm91$ & $4.10\pm0.20$ & $-0.50\pm0.14$ & $77.2\pm1.3$ & $1458\pm57$ & (2)\\
    4165030 & $5675\pm91$ & $4.30\pm0.17$ & $-0.29\pm0.10$ & $61.8\pm1.5$ & $1102\pm52$ & (1)\\
    4252818 & $5680\pm91$ & $4.30\pm0.17$ & $0.08\pm0.09$ & $69.2\pm7.4$ & $1318\pm65$ & (1)\\

    4349452 & $6161\pm102$ & $4.16\pm0.20$ & $-0.00\pm0.09$ & $98.27\pm0.57$ & $2106\pm50$ & (2)\\
    4465324 & $5767\pm111$ & $4.21\pm0.19$ & $0.03\pm0.10$ & $86.2\pm2.8$ & $1744\pm64$ & (1)\\
    4543171 & $5886\pm114$ & $4.23\pm0.19$ & $0.16\pm0.10$ & $71.0\pm1.7$ & $1480\pm54$ & (1)\\
    4554830 & $5509\pm86$ & $4.18\pm0.18$ & $0.32\pm0.09$ & $85.6\pm1.2$ & $1751\pm42$ & (1)\\

    4577484 & $5588\pm90$ & $4.23\pm0.18$ & $0.22\pm0.09$ & $47.0\pm1.4$ & $817\pm63$ & (1)\\
    4646780 & $6344\pm108$ & $4.11\pm0.21$ & $-0.14\pm0.10$ & $59.3\pm1.8$ & $1104\pm76$ & (1)\\
    4672403 & $5774\pm92$ & $4.04\pm0.21$ & $0.23\pm0.09$ & $60.3\pm1.2$ & $1107\pm44$ & (1)\\
    4739932 & $5847\pm94$ & $4.21\pm0.19$ & $0.18\pm0.09$ & $60.9\pm1.4$ & $1132\pm78$ & (1)\\
    4755204 & $5913\pm95$ & $4.24\pm0.19$ & $0.07\pm0.09$ & $70.4\pm1.2$ & $1360\pm42$ & (1)\\
    4841753 & $5976\pm97$ & $4.14\pm0.20$ & $0.19\pm0.09$ & $50.2\pm1.9$ & $871\pm103$ & (1)\\
    4842436 & $5761\pm92$ & $4.26\pm0.18$ & $0.24\pm0.09$ & $72.6\pm3.0$ & $1339\pm115$ & (1)\\
    4914423 & $5852\pm94$ & $4.23\pm0.18$ & $0.15\pm0.09$ & $81.5\pm1.6$ & $1663\pm56$ & (2)\\
    4914923 & $5766\pm92$ & $4.20\pm0.19$ & $0.21\pm0.09$ & $88.6\pm0.3$ & $1849\pm46$ & (1)\\
    4947253 & $5909\pm95$ & $4.12\pm0.20$ & $-0.02\pm0.09$ & $52.6\pm1.0$ & $923\pm40$ & (1)\\
    5088536 & $5830\pm93$ & $4.23\pm0.18$ & $-0.11\pm0.09$ & $109.6\pm3.1$ & $   $ & (2)\\
    5094751 & $5826\pm93$ & $4.24\pm0.19$ & $0.01\pm0.09$ & $91.1\pm2.3$ & $1745\pm117$ & (2)\\
    5095159 & $5305\pm87$ & $4.24\pm0.17$ & $0.18\pm0.09$ & $35.0\pm0.5$ & $582\pm29$ & (1)\\
    5095850 & $6622\pm118$ & $4.00\pm0.23$ & $0.15\pm0.11$ & $46.7\pm3.2$ & $791\pm37$ & (1)\\
    5253542 & $5685\pm114$ & $4.22\pm0.19$ & $0.25\pm0.10$ & $108.4\pm1.8$ & $   $ & (1)\\

    5353186 & $6038\pm98$ & $4.12\pm0.20$ & $0.17\pm0.09$ & $47.4\pm1.2$ & $872\pm31$ & (1)\\

    5511081 & $5826\pm93$ & $4.25\pm0.18$ & $0.02\pm0.09$ & $63.3\pm3.4$ & $   $ & (2)\\
    5512589 & $5720\pm91$ & $4.15\pm0.19$ & $0.20\pm0.09$ & $68.2\pm0.7$ & $1224\pm43$ & (1)\\

    5523099 & $5513\pm89$ & $4.24\pm0.18$ & $0.03\pm0.09$ & $41.8\pm0.9$ & $722\pm31$ & (1)\\

    5561278 & $5984\pm97$ & $4.00\pm0.15$ & $-0.03\pm0.09$ & $56.2\pm1.7$ & $1023\pm36$ & (2)\\
    5636956 & $6214\pm103$ & $4.10\pm0.21$ & $0.21\pm0.10$ & $55.4\pm0.9$ & $1007\pm36$ & (1)\\
    5686856 & $5897\pm95$ & $4.16\pm0.19$ & $0.11\pm0.09$ & $62.3\pm1.7$ & $1081\pm29$ & (1)\\
    5689219 & $6278\pm105$ & $4.16\pm0.20$ & $-0.24\pm0.11$ & $64.8\pm1.0$ & $1282\pm76$ & (1)\\
    5780885 & $5969\pm97$ & $4.08\pm0.21$ & $0.19\pm0.09$ & $56.4\pm1.7$ & $   $ & (2)\\

    5961597 & $6300\pm106$ & $4.09\pm0.21$ & $0.14\pm0.10$ & $59.8\pm0.9$ & $1258\pm43$ & (1)\\
    6064910 & $6172\pm102$ & $4.03\pm0.22$ & $-0.21\pm0.10$ & $43.9\pm0.8$ & $721\pm43$ & (1)\\

    6196457 & $5877\pm94$ & $4.12\pm0.20$ & $0.19\pm0.09$ & $66.6\pm1.1$ & $1299\pm53$ & (2)\\
    6268648 & $6032\pm98$ & $4.09\pm0.21$ & $-0.25\pm0.10$ & $88.9\pm2.0$ & $   $ & (2)\\
    6308642 & $5671\pm91$ & $4.17\pm0.19$ & $-0.13\pm0.09$ & $42.5\pm0.7$ & $733\pm28$ & (1)\\
    6520835 & $6029\pm98$ & $4.02\pm0.22$ & $-0.01\pm0.09$ & $49.7\pm0.8$ & $890\pm32$ & (1)\\
    6521045 & $5806\pm93$ & $4.06\pm0.21$ & $0.12\pm0.09$ & $77.0\pm1.1$ & $1502\pm31$ & (2)\\
    6587236 & $5921\pm95$ & $4.08\pm0.21$ & $-0.23\pm0.10$ & $32.1\pm1.9$ & $481\pm25$ & (1)\\
    6592305 & $6001\pm97$ & $4.13\pm0.20$ & $0.14\pm0.09$ & $46.8\pm0.6$ & $842\pm23$ & (1)\\
    6593461 & $5658\pm90$ & $4.23\pm0.18$ & $0.23\pm0.09$ & $90.8\pm2.0$ & $1927\pm338$ & (1)\\
    6603624 & $5515\pm89$ & $4.11\pm0.19$ & $0.29\pm0.09$ & $110.4\pm1.7$ & $2402\pm51$ & (1)\\
    6605673 & $6007\pm98$ & $4.02\pm0.22$ & $-0.26\pm0.11$ & $68.0\pm0.9$ & $1273\pm49$ & (1)\\
    6688822 & $5559\pm89$ & $4.05\pm0.20$ & $0.29\pm0.09$ & $47.1\pm1.0$ & $811\pm33$ & (1)\\
    6689943 & $5942\pm96$ & $4.14\pm0.20$ & $0.10\pm0.09$ & $80.7\pm1.5$ & $1682\pm66$ & (1)\\
    6693861 & $5749\pm92$ & $4.18\pm0.19$ & $-0.28\pm0.10$ & $46.7\pm1.0$ & $765\pm42$ & (1)\\
    6766513 & $6154\pm102$ & $4.17\pm0.20$ & $-0.06\pm0.10$ & $51.3\pm1.1$ & $883\pm84$ & (1)\\

    6853020 & $6161\pm102$ & $4.02\pm0.22$ & $0.15\pm0.10$ & $54.8\pm1.3$ & $   $ & (1)\\
    6863041 & $5622\pm90$ & $4.06\pm0.20$ & $0.24\pm0.09$ & $41.9\pm1.1$ & $775\pm32$ & (1)\\

    7038145 & $5896\pm115$ & $4.02\pm0.22$ & $0.06\pm0.10$ & $43.0\pm0.7$ & $764\pm30$ & (1)\\

    7107778 & $5149\pm87$ & $4.07\pm0.19$ & $0.11\pm0.10$ & $31.4\pm0.6$ & $529\pm15$ & (1)\\
    7133688 & $6291\pm106$ & $4.08\pm0.21$ & $0.17\pm0.10$ & $59.7\pm2.1$ & $1146\pm74$ & (1)\\
    7199397 & $5915\pm95$ & $4.02\pm0.21$ & $-0.07\pm0.09$ & $38.6\pm0.68$ & $643\pm17$ & (2)\\
    7264595 & $5665\pm91$ & $4.01\pm0.21$ & $-0.12\pm0.09$ & $36.3\pm1.6$ & $543\pm24$ & (1)\\
    7282890 & $6266\pm105$ & $4.13\pm0.20$ & $0.22\pm0.10$ & $46.0\pm0.7$ & $834\pm49$ & (1)\\
    7383120 & $5995\pm97$ & $4.20\pm0.19$ & $-0.08\pm0.09$ & $85.6\pm2.5$ & $1794\pm85$ & (1)\\
    7386523 & $5840\pm94$ & $4.11\pm0.20$ & $0.15\pm0.09$ & $48.3\pm2.0$ & $919\pm113$ & (1)\\
    7429287 & $5618\pm90$ & $4.22\pm0.19$ & $-0.31\pm0.11$ & $71.2\pm1.4$ & $1345\pm34$ & (1)\\

    7591963 & $5903\pm95$ & $4.19\pm0.19$ & $0.13\pm0.09$ & $59.2\pm0.9$ & $1096\pm48$ & (1)\\

    7680114 & $5810\pm93$ & $4.23\pm0.18$ & $0.18\pm0.09$ & $85.1\pm1.3$ & $1684\pm47$ & (1)\\

    7833440 & $6051\pm99$ & $4.13\pm0.20$ & $-0.27\pm0.11$ & $66.2\pm1.1$ & $1118\pm80$ & (1)\\
    7880676 & $6019\pm98$ & $4.14\pm0.20$ & $0.11\pm0.09$ & $53.7\pm1.0$ & $1017\pm43$ & (1)\\
    7910848 & $5964\pm96$ & $4.23\pm0.19$ & $0.16\pm0.09$ & $73.3\pm1.4$ & $   $ & (1)\\

    8012842 & $5712\pm91$ & $4.26\pm0.18$ & $0.22\pm0.09$ & $95.8\pm2.3$ & $2000\pm251$ & (1)\\
    8016496 & $6009\pm98$ & $3.97\pm0.22$ & $-0.03\pm0.09$ & $55.0\pm1.2$ & $1053\pm48$ & (1)\\
    8019508 & $5994\pm97$ & $4.08\pm0.21$ & $0.14\pm0.09$ & $37.7\pm2.0$ & $658\pm35$ & (1)\\
    8045442 & $5899\pm95$ & $4.08\pm0.21$ & $0.15\pm0.09$ & $31.8\pm0.6$ & $   $ & (1)\\

    8298626 & $5969\pm97$ & $4.13\pm0.20$ & $-0.02\pm0.09$ & $91.4\pm1.6$ & $1861\pm69$ & (1)\\

    8349582 & $5600\pm90$ & $4.10\pm0.20$ & $0.24\pm0.09$ & $83.6\pm1.4$ & $1677\pm90$ & (2)\\
    8367710 & $6117\pm140$ & $4.12\pm0.22$ & $0.07\pm0.11$ & $56.1\pm1.1$ & $1085\pm52$ & (1)\\
    8391021 & $6094\pm100$ & $4.06\pm0.21$ & $-0.29\pm0.11$ & $79.7\pm1.4$ & $1595\pm45$ & (1)\\
    8394589 & $6011\pm98$ & $4.11\pm0.20$ & $-0.28\pm0.11$ & $109.5\pm1.9$ & $2165\pm124$ & (1)\\
    8420801 & $6155\pm102$ & $4.17\pm0.20$ & $0.13\pm0.10$ & $67.4\pm1.8$ & $1287\pm86$ & (1)\\

    8491374 & $6197\pm103$ & $4.17\pm0.20$ & $0.09\pm0.10$ & $57.5\pm2.2$ & $1022\pm94$ & (1)\\
    8493735 & $5884\pm95$ & $4.02\pm0.21$ & $-0.02\pm0.09$ & $38.86\pm1.94$ & $586.1\pm29.3$ & (3)\\
    8493800 & $5921\pm95$ & $4.15\pm0.20$ & $-0.00\pm0.09$ & $83.6\pm1.8$ & $1850\pm109$ & (1)\\
    8494142 & $5955\pm96$ & $4.03\pm0.22$ & $-0.04\pm0.09$ & $61.8\pm0.76$ & $1133\pm81$ & (2)\\

    8554498 & $5866\pm94$ & $4.21\pm0.19$ & $0.19\pm0.09$ & $61.98\pm0.96$ & $1153\pm76$ & (2)\\
    8621637 & $5679\pm91$ & $4.22\pm0.18$ & $0.18\pm0.09$ & $57.1\pm1.4$ & $1032\pm47$ & (1)\\
    8684730 & $5915\pm95$ & $4.13\pm0.20$ & $0.15\pm0.09$ & $51.7\pm1.9$ & $962\pm39$ & (2)\\

    8776961 & $5816\pm93$ & $4.21\pm0.09$ & $0.08\pm0.09$ & $64.7\pm1.3$ & $1208\pm33$ & (1)\\
    8802782 & $5864\pm94$ & $3.97\pm0.22$ & $0.24\pm0.09$ & $44.9\pm1.7$ & $784\pm54$ & (1)\\
    8817551 & $5763\pm92$ & $4.02\pm0.21$ & $0.25\pm0.09$ & $42.7\pm0.8$ & $740\pm34$ & (1)\\

    8868481 & $5591\pm90$ & $4.12\pm0.20$ & $0.04\pm0.09$ & $40.4\pm1.3$ & $673\pm42$ & (1)\\
    8915084 & $5814\pm93$ & $4.06\pm0.21$ & $0.13\pm0.09$ & $79.1\pm1.4$ & $1594\pm59$ & (1)\\
    8938364 & $5639\pm90$ & $4.31\pm0.17$ & $-0.14\pm0.09$ & $85.8\pm1.1$ & $1681\pm37$ & (1)\\
    8956017 & $6221\pm104$ & $4.00\pm0.22$ & $0.15\pm0.10$ & $62.2\pm2.2$ & $1234\pm51$ & (1)\\
    8981766 & $6190\pm103$ & $4.14\pm0.20$ & $0.19\pm0.10$ & $63.1\pm1.6$ & $1277\pm32$ & (1)\\
    9005973 & $5821\pm93$ & $4.28\pm0.18$ & $0.00\pm0.09$ & $82.6\pm2$ & $1560\pm141$ & (1)\\

    9116461 & $6222\pm104$ & $4.17\pm0.20$ & $0.03\pm0.10$ & $104.7\pm2.3$ & $2334\pm197$ & (1)\\

    9335972 & $5719\pm91$ & $4.02\pm0.21$ & $0.13\pm0.09$ & $46.2\pm1.6$ & $802\pm20$ & (1)\\

    9446628 & $6146\pm101$ & $4.17\pm0.20$ & $-0.04\pm0.09$ & $56.1\pm2.5$ & $1031\pm76$ & (1)\\
    9451706 & $5944\pm96$ & $4.15\pm0.20$ & $0.22\pm0.09$ & $95.0\pm1.6$ & $1988\pm86$ & (2)\\
    9451741 & $5666\pm91$ & $4.20\pm0.19$ & $0.22\pm0.09$ & $94.4\pm1.9$ & $2079\pm105$ & (1)\\

    9592705 & $6049\pm99$ & $4.14\pm0.20$ & $0.17\pm0.09$ & $53.5\pm0.32$ & $1008\pm21$ & (2)\\
    9664694 & $6160\pm102$ & $4.17\pm0.20$ & $-0.01\pm0.09$ & $41.2\pm1.1$ & $722\pm27$ & (1)\\

    9696358 & $5987\pm97$ & $4.04\pm0.21$ & $0.12\pm0.09$ & $51.4\pm3.7$ & $    $ & (2)\\
    9697131 & $6143\pm101$ & $4.11\pm0.21$ & $0.04\pm0.09$ & $60.2\pm1.2$ & $1196\pm80$ & (1)\\
    9700430 & $5883\pm95$ & $4.24\pm0.18$ & $0.13\pm0.09$ & $78.9\pm1.5$ & $1623\pm94$ & (1)\\
    9715099 & $6133\pm101$ & $4.07\pm0.21$ & $0.15\pm0.09$ & $40.7\pm0.7$ & $   $ & (1)\\
    9754284 & $5941\pm96$ & $4.19\pm0.19$ & $0.09\pm0.09$ & $73.8\pm1.5$ & $1496\pm90$ & (1)\\
    9757640 & $5505\pm89$ & $4.17\pm0.19$ & $0.30\pm0.09$ & $61.7\pm1.0$ & $1142\pm86$ & (1)\\
    9778067 & $5899\pm95$ & $4.13\pm0.20$ & $-0.45\pm0.13$ & $50.0\pm1.7$ & $890\pm46$ & (1)\\
    9787965 & $6064\pm99$ & $4.16\pm0.20$ & $0.09\pm0.09$ & $53.4\pm2.3$ & $912\pm60$ & (1)\\
    9791157 & $5845\pm94$ & $4.17\pm0.19$ & $0.12\pm0.09$ & $54.4\pm1.0$ & $984\pm43$ & (1)\\

    9872292 & $6148\pm101$ & $4.17\pm0.20$ & $-0.01\pm0.09$ & $63.8\pm1.2$ & $   $ & (2)\\

    9962623 & $5684\pm91$ & $4.19\pm0.19$ & $-0.00\pm0.09$ & $82.2\pm2.5$ & $1543\pm164$ & (1)\\

    10019747 & $5572\pm89$ & $4.15\pm0.19$ & $0.28\pm0.09$ & $66.7\pm1.5$ & $1248\pm41$ & (1)\\

    10079226 & $5849\pm94$ & $4.24\pm0.18$ & $0.18\pm0.09$ & $116.4\pm1.9$ & $2689\pm93$ & (1)\\

    10322381 & $6056\pm99$ & $4.21\pm0.19$ & $-0.26\pm0.11$ & $86.6\pm4.3$ & $1657\pm155$ & (1)\\
    10351059 & $6247\pm104$ & $4.09\pm0.21$ & $0.23\pm0.10$ & $65.0\pm0.9$ & $1329\pm68$ & (1)\\
    10398597 & $6153\pm102$ & $4.10\pm0.21$ & $0.19\pm0.10$ & $57.6\pm1.5$ & $973\pm85$ & (1)\\
    10417911 & $5556\pm89$ & $4.06\pm0.20$ & $0.29\pm0.09$ & $56.1\pm2.8$ & $1022\pm61$ & (1)\\
    10586004 & $5766\pm92$ & $4.19\pm0.19$ & $0.25\pm0.09$ & $69.2\pm1.4$ & $1395\pm40$ & (2)\\

    10727922 & $5997\pm97$ & $4.11\pm0.20$ & $-0.15\pm0.10$ & $55.4\pm2.9$ & $979\pm56$ & (1)\\
    10731424 & $6158\pm102$ & $4.02\pm0.22$ & $0.03\pm0.09$ & $37.6\pm2.3$ & $670\pm45$ & (1)\\
    10732098 & $5766\pm92$ & $4.19\pm0.19$ & $0.13\pm0.09$ & $62.1\pm1.1$ & $1082\pm37$ & (1)\\
    10875245 & $5707\pm91$ & $4.22\pm0.18$ & $0.25\pm0.09$ & $86.7\pm3.7$ & $1711\pm107$ & (2)\\
    10923629 & $6013\pm98$ & $4.05\pm0.21$ & $0.20\pm0.09$ & $42.5\pm0.8$ & $711\pm37$ & (1)\\

    11083308 & $6021\pm98$ & $4.20\pm0.19$ & $0.07\pm0.09$ & $51.2\pm1.3$ & $902\pm39$ & (1)\\
    11133306 & $5910\pm95$ & $4.23\pm0.18$ & $0.00\pm0.09$ & $107.9\pm1.9$ & $2381\pm95$ & (2)\\

    11138101 & $6152\pm102$ & $4.10\pm0.21$ & $0.05\pm0.09$ & $43.8\pm1.3$ & $   $ & (1)\\
    11188219 & $5793\pm93$ & $4.26\pm0.18$ & $0.24\pm0.09$ & $95.6\pm3.4$ & $1843\pm106$ & (1)\\

    11193681 & $5597\pm90$ & $4.05\pm0.20$ & $0.26\pm0.09$ & $42.1\pm0.7$ & $749\pm36$ & (1)\\
    11244118 & $5600\pm90$ & $4.11\pm0.20$ & $0.26\pm0.09$ & $71.3\pm0.9$ & $1420\pm31$ & (1)\\

    11506988 & $6220\pm104$ & $4.21\pm0.19$ & $-0.16\pm0.10$ & $58.7\pm1.0$ & $1056\pm81$ & (1)\\
    11507653 & $5847\pm94$ & $4.17\pm0.19$ & $0.22\pm0.09$ & $70.0\pm2.3$ & $1359\pm217$ & (1)\\

    11611414 & $6033\pm98$ & $4.20\pm0.19$ & $-0.02\pm0.09$ & $68.3\pm3.3$ & $1162\pm127$ & (1)\\

    11771760 & $5859\pm94$ & $4.09\pm0.20$ & $0.04\pm0.09$ & $32.4\pm0.7$ & $535\pm19$ & (1)\\

    11817562 & $5970\pm97$ & $4.01\pm0.22$ & $0.20\pm0.09$ & $41.1\pm0.7$ & $713\pm46$ & (1)\\
    11919192 & $6358\pm108$ & $4.02\pm0.22$ & $-0.16\pm0.11$ & $46.8\pm1.7$ & $903\pm48$ & (1)\\

    11971746 & $5861\pm94$ & $4.22\pm0.19$ & $0.25\pm0.09$ & $90.8\pm2.1$ & $1911\pm124$ & (1)\\
    12068975 & $5958\pm96$ & $4.17\pm0.19$ & $-0.31\pm0.11$ & $108.4\pm3.1$ & $2298\pm105$ & (2)\\
    12069127 & $6223\pm104$ & $4.13\pm0.20$ & $0.18\pm0.10$ & $48.2\pm0.9$ & $829\pm41$ & (1)\\

    12265063 & $5944\pm96$ & $4.22\pm0.19$ & $-0.05\pm0.09$ & $67.9\pm1.5$ & $1314\pm139$ & (1)\\
    \emph{4167879} & $5770\pm109$ & $4.14\pm0.20$ & $-0.24\pm0.11$ & $3.22\pm0.059$ & $34.48\pm5.64$ & (6)\\
    4245297 & $5763\pm92$ & $4.22\pm0.18$ & $0.16\pm0.09$ & $3.42\pm0.078$ & $32.50\pm3.35$ & (6)\\
    4466582 & $5627\pm90$ & $4.30\pm0.17$ & $0.09\pm0.09$ & $10.334\pm1.033$ & $116.809\pm11.68$ & (5)\\
    4581415 & $5827\pm103$ & $4.01\pm0.22$ & $-0.32\pm0.12$ & $3.5325\pm0.086$ & $30.454\pm2.157$ & (6)\\
    5176520 & $5444\pm88$ & $4.36\pm0.16$ & $-0.50\pm0.14$ & $12.81\pm1.281$ & $153.731\pm15.37$ & (5)\\
    5649546 & $5744\pm107$ & $4.06\pm0.21$ & $0.11\pm0.10$ & $  $ & $  $ & $  $\\
    5683538 & $5860\pm94$ & $4.09\pm0.20$ & $-0.52\pm0.14$ & $42\pm1.4$ & $695\pm32$ & (1)\\
    5814204 & $6288\pm106$ & $4.12\pm0.20$ & $0.03\pm0.10$ & $5.02\pm0.12$ & $51.924\pm2.37$ & (6)\\
    5892169 & $5730\pm88$ & $4.08\pm0.20$ & $0.08\pm0.09$ & $29.9\pm0.4$ & $492\pm16$ & (1)\\
    6209592 & $5946\pm145$ & $3.99\pm0.26$ & $-0.40\pm0.16$ & $  $ & $  $ & $  $\\
    6290627 & $5352\pm87$ & $4.06\pm0.20$ & $-0.20\pm0.10$ & $22.4\pm2.24$ & $318.4\pm31.84$ & (3)\\
    6347797 & $6664\pm136$ & $4.15\pm0.20$ & $-0.24\pm0.13$ & $  $ & $   $ & $  $\\
    6363365 & $5894\pm95$ & $4.20\pm0.19$ & $-0.64\pm0.18$ & $86.3\pm1.4$ & $1629\pm80$ & (1)\\
    6763633 & $6321\pm107$ & $4.11\pm0.21$ & $0.20\pm0.10$ & $  $ & $  $ & $  $\\
    6852624 & $5700\pm92$ & $4.23\pm0.18$ & $-0.30\pm0.11$ & $  $ & $  $ & $  $\\
    7955597 & $5780\pm92$ & $4.15\pm0.19$ & $0.11\pm0.09$ & $12.46\pm0.23$ & $155.63\pm7.33$ & (6)\\
    8520552 & $7383\pm154$ & $4.02\pm0.22$ & $-0.57\pm0.22$ & $6.2147\pm0.172$ & $60.498\pm1.439$ & $(6)$\\
    8747199 & $5664\pm122$ & $4.25\pm0.19$ & $-0.60\pm0.21$ & $  $ & $  $ & $  $\\
    9697262 & $5409\pm88$ & $4.06\pm0.20$ & $-0.08\pm0.09$ & $  $ & $  $ & $  $\\
    9723621 & $5795\pm93$ & $4.10\pm0.20$ & $0.19\pm0.09$ & $6.04\pm0.13$ & $67.96\pm4.41$ & (6)\\
    9768011 & $7322\pm210$ & $3.96\pm0.25$ & $-0.80\pm0.35$ & $57\pm3.2$ & $918\pm48$ & (1)\\
    9778260 & $5543\pm89$ & $4.24\pm0.18$ & $0.01\pm0.09$ & $  $ & $  $ & $  $\\
    9906321 & $5604\pm90$ & $4.13\pm0.19$ & $-0.17\pm0.10$ & $5.24\pm0.15$ & $49.76\pm1.45$ & (6)\\
    10097375 & $5612\pm90$ & $4.32\pm0.17$ & $0.10\pm0.09$ & $  $ & $  $ & $  $\\
    10752443 & $6646\pm165$ & $4.06\pm0.23$ & $-0.29\pm0.15$ & $  $ & $  $ & $  $\\
    10873176 & $6058\pm99$ & $4.05\pm0.21$ & $-0.68\pm0.20$ & $49.6\pm0.9$ & $821\pm27$ & (1)\\
    10919121 & $5577\pm117$ & $4.16\pm0.21$ & $0.22\pm0.11$ & $  $ & $  $ & $  $\\
    11136690 & $6032\pm136$ & $4.16\pm0.21$ & $-0.81\pm0.31$ & $  $ & $  $ & $  $\\
    11466403 & $5598\pm90$ & $4.16\pm0.19$ & $-0.22\pm0.10$ & $8.024\pm0.80$ & $84.626\pm8.46$ & (5)\\

 \end{longtable}
\end{center}
\begin{center}
\begin{list}{}{}
\item[$^{\rm{}}$] Reference: (1) \cite{Chaplin14}; (2) \cite{Huber13} (3)
\cite{Mosser12}; (4) \cite{Appourchaux}; (5) \cite{Stello13}; (6) \cite{Hekker11}
\end{list}
\end{center}

\clearpage

\begin{table}
\begin{center}
\caption{Input parameters for Grid Calculation.}\label{tbl3}
\begin{tabular}{cccc}
    \hline 
    \hline
    $[\rm{Fe/H}]_{\rm{ini}}$ (dex) & -0.3 $\sim$ +0.4 & $\delta [\rm{Fe/H}]$ (dex) & 0.1  \\
    $Z_{\rm ini}$ (dex) & 0.0085 $\sim$ 0.0400 &                             \\
    $M$ ($\Msun$) & 0.8 $\sim$ 2.5 & $\delta M$ ($\Msun$) & 0.02 \\
    $\alpha$ & 1.75, 1.842, 1.95 & $\delta \alpha$ & 0.1  \\
    \hline
    \hline
\end{tabular}
\end{center}
\end{table}

\clearpage

\begin{center}
\begin{longtable}{cccccccc}
    \caption{Fundamental Parameters of 150 stars. }\label{table:3}\\

    \endfirsthead																																
\multicolumn{8}{c}%
{{\tablename\ \thetable{} -- continued from previous page}} \\ \hline																																						
  Star & ${T_{\rm{eff}}}$ & $\log g$ & $[\rm{Fe/H}]$ & $M$ & $R$ & Age & $L$\\
    KIC & (K) & (dex) & (dex) & ($\Msun$) & ($\Rsun$) & (Gyr) & ($\Lsun$)\\
\hline																																						
\hline																																						
\endhead																																						
\hline \multicolumn{8}{c}{{Continued on next page}} \\																																						
\endfoot																																						
\hline																																						
\endlastfoot																																																																								\hline
  Star & ${T_{\rm{eff}}}$ & $\log g$ & $[\rm{Fe/H}]$ & $M$ & $R$ & Age & $L$\\
    KIC & (K) & (dex) & (dex) & ($\Msun$) & ($\Rsun$) & (Gyr) & ($\Lsun$) \\
				\hline																																		
				\hline	
  \hline
  \it{MSTO stars}&&&&&&&\\			
  \hline
   1725815 & $6225_{-93}^{+77}$ & $3.966_{-0.011}^{+0.008}$ & $0.01_{-0.10}^{+0.10}$ & $1.28_{-0.04}^{+0.04}$ & $1.95_{-0.03}^{+0.03}$ & $3.94_{-0.34}^{+0.55}$ & $5.14_{-0.38}^{0.36}$\\
   2865774 & $5807_{-60}^{+52}$ & $4.022_{-0.017}^{+0.017}$ & $0.10_{-0.08}^{+0.03}$ & $1.12_{-0.04}^{+0.04}$ & $1.71_{-0.04}^{+0.04}$ & $7.03_{-0.76}^{+0.69}$ & $2.99_{-0.16}^{0.19}$\\
   2998253 & $6107_{-56}^{+62}$ & $4.214_{-0.016}^{+0.018}$ & $0.10_{-0.11}^{+0.10}$ & $1.16_{-0.04}^{+0.02}$ & $1.39_{-0.04}^{+0.03}$ & $5.01_{-0.61}^{+0.81}$ & $2.42_{-0.15}^{0.17}$\\
   3112152 & $5980_{-62}^{+53}$ & $4.029_{-0.013}^{+0.013}$ & $-0.02_{-0.07}^{+0.02}$ & $1.14_{-0.04}^{+0.04}$ & $1.71_{-0.03}^{+0.03}$ & $6.04_{-0.66}^{+0.74}$ & $3.37_{-0.20}^{0.20}$\\
   3123191 & $6134_{-54}^{+75}$ & $4.217_{-0.012}^{+0.012}$ & $-0.02_{-0.10}^{+0.03}$ & $1.12_{-0.04}^{+0.02}$ & $1.37_{-0.02}^{+0.02}$ & $5.16_{-0.66}^{+0.61}$ & $2.39_{-0.11}^{0.13}$\\
   3241581 & $5580_{-54}^{+53}$ & $4.386_{-0.008}^{+0.007}$ & $0.34_{-0.15}^{+0.11}$ & $1.02_{-0.04}^{+0.04}$ & $1.08_{-0.01}^{+0.02}$ & $6.39_{-1.34}^{+1.93}$ & $1.02_{-0.06}^{+0.05}$\\

    3656476 & $5543_{-48}^{+75}$ & $4.224_{-0.006}^{+0.004}$ & $0.32_{-0.10}^{+0.10}$ & $1.04_{-0.04}^{+0.02}$ & $1.31_{-0.01}^{+0.01}$ & $9.69_{-1.59}^{+1.10}$ & $1.46_{-0.08}^{0.09}$\\

    3967859 & $5890_{-57}^{+56}$ & $4.218_{-0.015}^{+0.015}$ & $-0.24_{-0.10}^{+0.01}$ & $0.94_{-0.04}^{+0.04}$ & $1.25_{-0.03}^{+0.03}$ & $9.76_{-1.10}^{+1.90}$ & $1.70_{-0.11}^{0.12}$\\
    4141376 & $5898_{-53}^{+58}$ & $4.404_{-0.007}^{+0.006}$ & $-0.29_{-0.010}^{+0.11}$ & $0.92_{-0.02}^{+0.04}$ & $1.01_{-0.01}^{+0.01}$ & $6.66_{-1.37}^{+1.71}$ & $1.11_{-0.06}^{0.07}$\\

   4252818 & $5682_{-57}^{+55}$ & $4.065_{-0.019}^{+0.019}$ & $0.11_{-0.10}^{+0.10}$ & $1.06_{-0.04}^{+0.02}$ & $1.58_{-0.05}^{+0.05}$ & $8.75_{-0.79}^{+1.23}$ & $2.35_{-0.17}^{0.19}$\\

   4465324 & $5776_{-71}^{+70}$ & $4.188_{-0.013}^{+0.012}$ & $0.09_{-0.10}^{+0.10}$ & $1.04_{-0.04}^{+0.04}$ & $1.36_{-0.03}^{+0.03}$ & $8.44_{-1.19}^{+1.61}$ & $1.85_{-0.13}^{0.14}$\\
   4543171 & $5896_{-68}^{+61}$ & $4.101_{-0.009}^{+0.008}$ & $0.20_{-0.10}^{+0.10}$ & $1.16_{-0.04}^{+0.02}$ & $1.59_{-0.02}^{+0.02}$ & $6.05_{-0.58}^{+0.91}$ & $2.74_{-0.17}^{0.15}$\\
    4554830 & $5510_{-51}^{+49}$ & $4.180_{-0.007}^{+0.007}$ & $0.33_{-0.01}^{+0.01}$ & $1.02_{-0.02}^{+0.04}$ & $1.37_{-0.02}^{+0.02}$ & $10.26_{-1.01}^{+1.06}$ & $1.56_{-0.08}^{0.08}$\\
   4646780 & $6360_{-73}^{+59}$ & $3.994_{-0.014}^{+0.014}$ & $-0.20_{-0.10}^{+0.10}$ & $1.20_{-0.04}^{+0.04}$ & $1.84_{-0.04}^{+0.04}$ & $4.22_{-0.47}^{+0.47}$ & $4.97_{-0.35}^{0.30}$\\
    4739932 & $5837_{-55}^{+59}$ & $4.003_{-0.012}^{+0.011}$ & $0.13_{-0.04}^{+0.07}$ & $1.16_{-0.02}^{+0.04}$ & $1.79_{-0.03}^{+0.03}$ & $6.25_{-0.47}^{+0.60}$ & $3.34_{-0.17}^{0.17}$\\
   4755204 & $5925_{-58}^{+54}$ & $4.082_{-0.008}^{+0.009}$ & $0.10_{-0.10}^{+0.10}$ & $1.14_{-0.04}^{+0.04}$ & $1.61_{-0.02}^{+0.02}$ & $6.20_{-0.63}^{+1.02}$ & $2.87_{-0.15}^{0.16}$\\
   4842436 & $5779_{-59}^{+48}$ & $4.091_{-0.016}^{+0.018}$ & $0.20_{-0.10}^{+0.10}$ & $1.12_{-0.02}^{+0.02}$ & $1.58_{-0.02}^{+0.03}$ & $7.45_{-0.95}^{+0.73}$ & $2.50_{-0.14}^{0.15}$\\
   4914423 & $5860_{-54}^{+55}$ & $4.166_{-0.010}^{+0.008}$ & $0.19_{-0.10}^{+0.01}$ & $1.12_{-0.04}^{+0.02}$ & $1.45_{-0.02}^{+0.02}$ & $6.52_{-0.63}^{+0.98}$ & $2.24_{-0.11}^{0.11}$\\
    4914923 & $5785_{-50}^{+44}$ & $4.208_{-0.004}^{+0.004}$ & $0.20_{-0.02}^{+0.01}$ & $1.10_{-0.04}^{+0.02}$ & $1.37_{-0.01}^{+0.01}$ & $7.11_{-0.80}^{+0.90}$ & $1.89_{-0.12}^{0.07}$\\
   4947253 & $5903_{-59}^{+57}$ & $3.917_{-0.010}^{+0.009}$ & $-0.06_{-0.04}^{+0.06}$ & $1.18_{-0.04}^{+0.02}$ & $1.97_{-0.03}^{+0.03}$ & $5.58_{-0.37}^{+0.45}$ & $4.25_{-0.24}^{0.24}$\\
    5094751 & $5833_{-56}^{+58}$ & $4.215_{-0.010}^{+0.011}$ & $-0.02_{-0.10}^{+0.10}$ & $1.04_{-0.04}^{+0.04}$ & $1.33_{-0.02}^{+0.02}$ & $7.77_{-1.22}^{+1.16}$ & $1.83_{-0.11}^{0.11}$\\
    5095850 & $6647_{-72}^{+66}$ & $3.883_{-0.016}^{+0.015}$ & $0.09_{-0.10}^{+0.10}$ & $1.60_{-0.04}^{+0.12}$ & $2.42_{-0.06}^{+0.07}$ & $1.84_{-0.42}^{+0.20}$ & $10.29_{-0.73}^{0.78}$\\

    5511081 & $5843_{-61}^{+50}$ & $4.013_{-0.023}^{+0.024}$ & $0.02_{-0.02}^{+0.07}$ & $1.10_{-0.02}^{+0.06}$ & $1.73_{-0.06}^{+0.05}$ & $7.04_{-0.92}^{+0.62}$ & $3.13_{-0.24}^{0.21}$\\
    5512589 & $5733_{-57}^{+51}$ & $4.055_{-0.006}^{+0.006}$ & $0.20_{-0.07}^{+0.02}$ & $1.10_{-0.02}^{+0.04}$ & $1.64_{-0.01}^{+0.02}$ & $7.80_{-0.72}^{+0.58}$ & $2.62_{-0.13}^{0.13}$\\

    5636956 & $6243_{-72}^{+50}$ & $3.969_{-0.008}^{+0.008}$ & $0.20_{-0.10}^{+0.08}$ & $1.36_{-0.02}^{+0.04}$ & $2.01_{-0.02}^{+0.03}$ & $3.25_{-0.26}^{+0.45}$ & $5.65_{-0.37}^{0.31}$\\
    5686856 & $5882_{-50}^{+63}$ & $3.998_{-0.007}^{+0.008}$ & $0.09_{-0.06}^{+0.04}$ & $1.16_{-0.04}^{+0.04}$ & $1.79_{-0.03}^{+0.02}$ & $6.26_{-0.53}^{+0.53}$ & $3.46_{-0.15}^{0.17}$\\
    5689219 & $6373_{-60}^{+51}$ & $4.039_{-0.008}^{+0.009}$ & $-0.29_{-0.04}^{+0.07}$ & $1.16_{-0.06}^{+0.02}$ & $1.71_{-0.03}^{+0.02}$ & $4.64_{-0.41}^{+0.85}$ & $4.06_{-0.20}^{0.17}$\\
    5780885 & $5987_{-69}^{+54}$ & $3.966_{-0.014}^{+0.015}$ & $0.20_{-0.07}^{+0.10}$ & $1.26_{-0.04}^{+0.02}$ & $1.93_{-0.04}^{+0.04}$ & $4.84_{-0.38}^{+0.40}$ & $4.31_{-0.28}^{0.25}$\\

    5961597 & $6338_{-62}^{+38}$ & $4.034_{-0.006}^{+0.005}$ & $0.20_{-0.10}^{+0.10}$ & $1.46_{-0.04}^{+0.02}$ & $1.92_{-0.02}^{+0.02}$ & $2.32_{-0.18}^{+0.22}$ & $5.29_{-0.16}^{0.21}$\\

    6196457 & $5865_{-51}^{+54}$ & $4.055_{-0.009}^{+0.009}$ & $0.20_{-0.08}^{+0.10}$ & $1.16_{-0.02}^{+0.02}$ & $1.69_{-0.02}^{+0.02}$ & $6.08_{-0.39}^{+0.81}$ & $3.03_{-0.16}^{0.22}$\\
    6268648 & $6033_{-60}^{+67}$ & $4.198_{-0.012}^{+0.012}$ & $-0.25_{-0.10}^{+0.10}$ & $1.00_{-0.04}^{+0.04}$ & $1.33_{-0.02}^{+0.02}$ & $7.46_{-1.02}^{+0.97}$ & $2.10_{-0.13}^{0.13}$\\
    6521045 & $5816_{-55}^{+50}$ & $4.127_{-0.007}^{+0.006}$ & $0.19_{-0.09}^{+0.01}$ & $1.10_{-0.02}^{+0.04}$ & $1.51_{-0.02}^{+0.02}$ & $7.18_{-0.90}^{+0.85}$ & $2.36_{-0.13}^{0.11}$\\
    6593461 & $5660_{-53}^{+48}$ & $4.216_{-0.012}^{+0.011}$ & $0.18_{-0.01}^{+0.01}$ & $1.04_{-0.02}^{+0.02}$ & $1.32_{-0.02}^{+0.02}$ & $8.86_{-1.33}^{+1.38}$ & $1.64_{-0.09}^{0.10}$\\
    6605673 & $6010_{-61}^{+58}$ & $4.051_{-0.007}^{+0.007}$ & $-0.24_{-0.09}^{+0.03}$ & $1.04_{-0.02}^{+0.04}$ & $1.61_{-0.02}^{+0.02}$ & $7.20_{-0.96}^{+0.70}$ & $3.03_{-0.17}^{0.17}$\\
    6689943 & $5947_{-53}^{+51}$ & $4.164_{-0.010}^{+0.008}$ & $0.10_{-0.01}^{+0.01}$ & $1.12_{-0.02}^{+0.04}$ & $1.46_{-0.02}^{+0.02}$ & $6.01_{-0.79}^{+0.75}$ & $2.42_{-0.11}^{0.11}$\\
    6853020 & $6171_{-64}^{+62}$ & $3.956_{-0.012}^{+0.012}$ & $0.10_{-0.10}^{+0.10}$ & $1.30_{-0.02}^{+0.04}$ & $2.00_{-0.04}^{+0.04}$ & $3.90_{-0.43}^{+0.31}$ & $5.20_{-0.29}^{0.29}$\\
    7133688 & $6314_{-73}^{+54}$ & $4.013_{-0.015}^{+0.016}$ & $0.10_{-0.10}^{+0.10}$ & $1.34_{-0.04}^{+0.08}$ & $1.90_{-0.05}^{+0.05}$ & $3.33_{-0.83}^{+0.46}$ & $5.16_{-0.42}^{0.36}$\\
    7383120 & $6106_{-76}^{+50}$ & $4.196_{-0.012}^{+0.011}$ & $-0.04_{-0.10}^{+0.10}$ & $1.08_{-0.02}^{+0.04}$ & $1.38_{-0.02}^{+0.03}$ & $6.08_{-0.87}^{+0.87}$ & $2.24_{-0.13}^{0.12}$\\
    7429287 & $5625_{-79}^{+55}$ & $4.059_{-0.007}^{+0.008}$ & $-0.29_{-0.09}^{+0.10}$ & $0.88_{-0.04}^{+0.02}$ & $1.45_{-0.02}^{+0.02}$ & $14.03_{-1.18}^{+0.37}$ & $1.89_{-0.10}^{0.11}$\\

    7591963 & $5888_{-53}^{+69}$ & $3.987_{-0.008}^{+0.008}$ & $0.12_{-0.03}^{+0.08}$ & $1.18_{-0.04}^{+0.02}$ & $1.83_{-0.02}^{+0.03}$ & $5.85_{-0.40}^{+0.53}$ & $3.62_{-0.16}^{0.22}$\\

    7680114 & $5815_{-55}^{+61}$ & $4.182_{-0.007}^{+0.007}$ & $0.19_{-0.10}^{+0.10}$ & $1.10_{-0.04}^{+0.02}$ & $1.41_{-0.02}^{+0.02}$ & $7.21_{-1.10}^{+0.99}$ & $2.05_{-0.12}^{0.12}$\\

    7833440 & $6050_{-61}^{+62}$ & $4.034_{-0.009}^{+0.009}$ & $-0.29_{-0.04}^{+0.08}$ & $1.06_{-0.02}^{+0.04}$ & $1.65_{-0.02}^{+0.02}$ & $6.70_{-0.85}^{+0.65}$ & $3.29_{-0.19}^{0.21}$\\
   7910848 & $5959_{-56}^{+58}$ & $4.106_{-0.010}^{+0.010}$ & $0.12_{-0.02}^{+0.08}$ & $1.16_{-0.04}^{+0.02}$ & $1.58_{-0.02}^{+0.02}$ & $5.84_{-0.51}^{+0.89}$ & $2.84_{-0.15}^{0.14}$\\
   8012842 & $5713_{-48}^{+80}$ & $4.250_{-0.011}^{+0.011}$ & $0.18_{-0.10}^{+0.10}$ & $1.06_{-0.02}^{+0.02}$ & $1.29_{-0.02}^{+0.02}$ & $7.60_{-1.08}^{+1.13}$ & $1.59_{-0.09}^{0.11}$\\

    8298626 & $5980_{-78}^{+51}$ & $4.224_{-0.009}^{+0.009}$ & $-0.02_{-0.10}^{+0.10}$ & $1.08_{-0.04}^{+0.02}$ & $1.33_{-0.02}^{+0.02}$ & $6.40_{-1.06}^{+0.96}$ & $2.02_{-0.11}^{0.12}$\\

    8349582 & $5602_{-52}^{+52}$ & $4.168_{-0.009}^{+0.009}$ & $0.19_{-0.10}^{+0.14}$ & $1.04_{-0.04}^{+0.04}$ & $1.40_{-0.02}^{+0.02}$ & $9.70_{-1.68}^{+1.10}$ & $1.73_{-0.10}^{0.09}$\\
    8367710 & $6151_{-102}^{+70}$ & $3.969_{-0.009}^{+0.010}$ & $0.02_{-0.10}^{+0.08}$ & $1.24_{-0.04}^{+0.04}$ & $1.92_{-0.03}^{+0.03}$ & $4.41_{-0.46}^{+0.70}$ & $4.75_{-0.40}^{+0.30}$\\
    8391021 & $6119_{-69}^{+51}$ & $4.150_{-0.008}^{+0.008}$ & $-0.27_{-0.09}^{+0.03}$ & $1.06_{-0.04}^{+0.04}$ & $1.44_{-0.02}^{+0.02}$ & $6.08_{-0.62}^{+1.15}$ & $2.60_{-0.16}^{+0.14}$\\
    8420801 & $6169_{-67}^{+57}$ & $4.069_{-0.013}^{+0.013}$ & $0.10_{-0.10}^{+0.10}$ & $1.24_{-0.04}^{+0.02}$ & $1.71_{-0.03}^{+0.03}$ & $4.47_{-0.43}^{+0.57}$ & $3.78_{-0.22}^{0.23}$\\

    8491374 & $6212_{-66}^{+56}$ & $3.978_{-0.017}^{+0.017}$ & $0.01_{-0.10}^{+0.10}$ & $1.26_{-0.02}^{+0.04}$ & $1.92_{-0.05}^{+0.05}$ & $4.11_{-0.32}^{+0.36}$ & $4.90_{-0.34}^{0.31}$\\

    8493800 & $5917_{-49}^{+59}$ & $4.184_{-0.008}^{+0.009}$ & $-0.02_{-0.01}^{+0.10}$ & $1.10_{-0.04}^{+0.02}$ & $1.40_{-0.02}^{+0.02}$ & $6.32_{-0.61}^{+1.17}$ & $2.19_{-0.11}^{0.10}$\\
    8494142 & $5968_{-67}^{+49}$ & $4.006_{-0.007}^{+0.007}$ & $-0.01_{-0.08}^{+0.02}$ & $1.14_{-0.04}^{+0.04}$ & $1.76_{-0.02}^{+0.02}$ & $6.07_{-0.82}^{+0.63}$ & $3.55_{-0.20}^{+0.18}$\\

    8554498 & $5849_{-52}^{+62}$ & $4.012_{-0.008}^{+0.008}$ & $0.14_{-0.02}^{+0.06}$ & $1.18_{-0.02}^{+0.02}$ & $1.78_{-0.02}^{+0.02}$ & $6.10_{-0.57}^{+0.58}$ & $3.32_{-0.15}^{0.18}$\\

    8776961 & $5832_{-61}^{+52}$ & $4.030_{-0.008}^{+0.008}$ & $0.09_{-0.08}^{+0.03}$ & $1.14_{-0.03}^{+0.03}$ & $1.70_{-0.03}^{+0.03}$ & $6.96_{-0.97}^{+0.70}$ & $3.00_{-0.15}^{0.16}$\\

    8915084 & $5828_{-58}^{+49}$ & $4.146_{-0.009}^{+0.009}$ & $0.19_{-0.10}^{+0.10}$ & $1.10_{-0.02}^{+0.04}$ & $1.48_{-0.02}^{+0.02}$ & $7.04_{-0.78}^{+0.88}$ & $2.27_{-0.14}^{0.11}$\\
    8938364 & $5639_{-57}^{+59}$ & $4.165_{-0.007}^{+0.006}$ & $-0.11_{-0.11}^{+0.01}$ & $0.92_{-0.04}^{+0.02}$ & $1.31_{-0.02}^{+0.02}$ & $13.27_{-1.20}^{+1.38}$ & $1.57_{-0.09}^{0.09}$\\
   8956017 & $6230_{-65}^{+61}$ & $4.041_{-0.012}^{+0.012}$ & $0.10_{-0.10}^{+0.10}$ & $1.28_{-0.04}^{+0.04}$ & $1.79_{-0.03}^{+0.04}$ & $3.93_{-0.47}^{+0.50}$ & $4.36_{-0.27}^{0.31}$\\
    8981766 & $6206_{-69}^{+57}$ & $4.056_{-0.009}^{+0.008}$ & $0.20_{-0.10}^{+0.10}$ & $1.30_{-0.04}^{+0.12}$ & $1.79_{-0.04}^{+0.06}$ & $3.73_{-1.20}^{+0.60}$ & $4.27_{-0.30}^{0.40}$\\
    9005973 & $5824_{-56}^{+63}$ & $4.160_{-0.012}^{+0.013}$ & $-0.01_{-0.01}^{+0.01}$ & $1.04_{-0.02}^{+0.04}$ & $1.41_{-0.03}^{+0.02}$ & $8.02_{-1.04}^{+1.40}$ & $2.07_{-0.13}^{0.13}$\\

    9446628 & $6147_{-67}^{+67}$ & $3.964_{-0.018}^{+0.018}$ & $-0.08_{-0.05}^{+0.08}$ & $1.20_{-0.04}^{+0.04}$ & $1.90_{-0.05}^{+0.05}$ & $4.83_{-0.52}^{+0.54}$ & $4.61_{-0.34}^{0.38}$\\
    9451706 & $5936_{-46}^{+63}$ & $4.252_{-0.008}^{+0.008}$ & $0.20_{-0.10}^{+0.10}$ & $1.12_{-0.02}^{+0.02}$ & $1.32_{-0.02}^{+0.02}$ & $5.90_{-0.79}^{+0.67}$ & $1.95_{-0.09}^{0.10}$\\
    9451741 & $5667_{-47}^{+83}$ & $4.243_{-0.010}^{+0.009}$ & $0.18_{-0.01}^{+0.01}$ & $1.04_{-0.02}^{+0.04}$ & $1.28_{-0.02}^{+0.02}$ & $8.09_{-0.92}^{+1.51}$ & $1.55_{-0.09}^{0.09}$\\

    9697131 & $6161_{-70}^{+56}$ & $4.005_{-0.009}^{+0.010}$ & $0.0_{-0.10}^{+0.10}$ & $1.22_{-0.04}^{+0.04}$ & $1.83_{-0.03}^{+0.03}$ & $4.61_{-0.38}^{+0.48}$ & $4.32_{-0.28}^{0.23}$\\
    9700430 & $5888_{-57}^{+59}$ & $4.147_{-0.010}^{+0.009}$ & $0.19_{-0.10}^{+0.01}$ & $1.12_{-0.02}^{+0.04}$ & $1.49_{-0.02}^{+0.02}$ & $6.43_{-0.59}^{+0.81}$ & $2.40_{-0.12}^{0.12}$\\
    9754284 & $5955_{-56}^{+50}$ & $4.111_{-0.011}^{+0.01}$ & $0.10_{-0.01}^{+0.01}$ & $1.14_{-0.04}^{+0.02}$ & $1.56_{-0.02}^{+0.02}$ & $5.99_{-0.59}^{+1.04}$ & $2.75_{-0.13}^{0.14}$\\

   9872292 & $6124_{-56}^{+84}$ & $4.030_{-0.010}^{+0.011}$ & $0.00_{-0.10}^{+0.10}$ & $1.18_{-0.02}^{+0.02}$ & $1.74_{-0.02}^{+0.02}$ & $5.17_{-0.37}^{+0.52}$ & $3.83_{-0.17}^{0.28}$\\
   9962623 & $5687_{-59}^{+58}$ & $4.148_{-0.015}^{+0.016}$ & $-0.01_{-0.01}^{+0.01}$ & $0.98_{-0.02}^{+0.04}$ & $1.39_{-0.02}^{+0.03}$ & $10.45_{-1.08}^{+1.20}$ & $1.82_{-0.11}^{0.11}$\\
   10019747 & $5591_{-63}^{+48}$ & $4.042_{-0.009}^{+0.009}$ & $0.23_{-0.01}^{+0.12}$ & $1.10_{-0.02}^{+0.04}$ & $1.67_{-0.03}^{+0.03}$ & $8.43_{-1.15}^{+0.72}$ & $2.44_{-0.14}^{0.12}$\\

   10322381 & $6063_{-60}^{+57}$ & $4.187_{-0.019}^{+0.019}$ & $-0.25_{-0.10}^{+0.01}$ & $1.02_{-0.04}^{+0.04}$ & $1.35_{-0.04}^{+0.04}$ & $7.00_{-0.90}^{+1.18}$ & $2.22_{-0.16}^{0.17}$\\
   10351059 & $6249_{-66}^{+65}$ & $4.061_{-0.009}^{+0.010}$ & $0.20_{-0.10}^{+0.10}$ & $1.32_{-0.04}^{+0.12}$ & $1.79_{-0.03}^{+0.05}$ & $3.48_{-1.07}^{+0.50}$ & $4.44_{-0.30}^{0.31}$\\
   10398597 & $6167_{-69}^{+59}$ & $3.981_{-0.012}^{+0.013}$ & $0.20_{-0.10}^{+0.10}$ & $1.32_{-0.06}^{+0.04}$ & $1.94_{-0.04}^{+0.04}$ & $3.90_{-0.44}^{+0.37}$ & $4.90_{-0.30}^{0.32}$\\
   10586004 & $5782_{-61}^{+52}$ & $4.079_{-0.008}^{+0.008}$ & $0.20_{-0.01}^{+0.10}$ & $1.12_{-0.02}^{+0.04}$ & $1.61_{-0.02}^{+0.02}$ & $7.18_{-0.83}^{+0.86}$ & $2.61_{-0.14}^{0.15}$\\

   10732098 & $5772_{-53}^{+53}$ & $4.000_{-0.007}^{+0.008}$ & $0.13_{-0.03}^{+0.07}$ & $1.14_{-0.04}^{+0.02}$ & $1.77_{-0.02}^{+0.02}$ & $7.10_{-0.87}^{+0.75}$ & $3.13_{-0.15}^{0.15}$\\
   10875245 & $5701_{-53}^{+66}$ & $4.189_{-0.015}^{+0.016}$ & $0.19_{-0.01}^{+0.14}$ & $1.08_{-0.04}^{+0.02}$ & $1.39_{-0.03}^{+0.03}$ & $7.79_{-0.65}^{+1.11}$ & $1.84_{-0.12}^{0.11}$\\

    11188219 & $5800_{-50}^{+47}$ & $4.239_{-0.009}^{+0.011}$ & $0.18_{-0.01}^{+0.02}$ & $1.10_{-0.04}^{+0.02}$ & $1.32_{-0.02}^{+0.02}$ & $6.81_{-0.73}^{+0.68}$ & $1.78_{-0.11}^{0.08}$\\

   11244118 & $5618_{-60}^{+50}$ & $4.087_{-0.006}^{+0.006}$ & $0.34_{-0.13}^{+0.10}$ & $1.10_{-0.04}^{+0.04}$ & $1.58_{-0.02}^{+0.02}$ & $8.06_{-1.17}^{+0.95}$ & $2.24_{-0.14}^{0.13}$\\

   11506988 & $6216_{-64}^{+65}$ & $3.979_{-0.009}^{+0.009}$ & $-0.20_{-0.1}^{+0.10}$ & $1.16_{-0.04}^{+0.04}$ & $1.83_{-0.03}^{+0.03}$ & $4.97_{-0.38}^{+0.45}$ & $4.49_{-0.25}^{0.26}$\\
   11507653 & $5854_{-59}^{+51}$ & $4.081_{-0.015}^{+0.015}$ & $0.20_{-0.01}^{+0.02}$ & $1.16_{-0.04}^{+0.02}$ & $1.62_{-0.04}^{+0.04}$ & $6.41_{-0.65}^{+1.03}$ & $2.79_{-0.17}^{0.16}$\\

   11611414 & $6039_{-59}^{+55}$ & $4.055_{-0.017}^{+0.018}$ & $-0.02_{-0.07}^{+0.02}$ & $1.14_{-0.04}^{+0.04}$ & $1.67_{-0.04}^{+0.04}$ & $5.84_{-0.72}^{+0.72}$ & $3.33_{-0.22}^{0.19}$\\

   11971746 & $5879_{-79}^{+45}$ & $4.228_{-0.011}^{+0.011}$ & $0.21_{-0.02}^{+0.08}$ & $1.13_{-0.05}^{+0.02}$ & $1.35_{-0.02}^{+0.02}$ & $6.44_{-0.67}^{+0.67}$ & $1.94_{-0.10}^{0.10}$\\

   12265063 & $5963_{-66}^{+51}$ & $4.058_{-0.011}^{+0.012}$ & $-0.02_{-0.07}^{+0.03}$ & $1.12_{-0.04}^{+0.04}$ & $1.64_{-0.03}^{+0.03}$ & $6.45_{-0.76}^{+0.73}$ & $3.05_{-0.18}^{0.18}$\\
																															
  \hline
  \it{MS stars}&&&&&&&\\			
  \hline		

3241581 & $5580_{-54}^{+53}$ & $4.386_{-0.008}^{+0.007}$ & $0.34_{-0.15}^{+0.02}$ & $1.02_{-0.04}^{+0.04}$ & $1.08_{-0.01}^{+0.02}$ & $6.39_{-1.34}^{+1.93}$ & $1.02_{-0.06}^{0.05}$\\
4349452 & $6159_{-58}^{+67}$ & $4.274_{-0.005}^{+0.005}$ & $-0.01_{-0.010}^{+0.01}$ & $1.14_{-0.04}^{+0.02}$ & $1.29_{-0.01}^{+0.01}$ & $4.10_{-0.85}^{+0.85}$ & $2.15_{-0.09}^{0.13}$\\
5088536 & $5818_{-49}^{+50}$ & $4.302_{-0.008}^{+0.011}$ & $-0.11_{-0.04}^{+0.05}$ & $0.94_{-0.02}^{+0.04}$ & $1.15_{-0.02}^{+0.01}$ & $9.68_{-1.17}^{+0.97}$ & $1.34_{-0.06}^{0.08}$\\
 5253542 & $5649_{-64}^{+73}$ & $4.313_{-0.009}^{+0.010}$ & $0.19_{-0.10}^{+0.15}$ & $1.02_{-0.04}^{+0.04}$ & $1.18_{-0.02}^{+0.02}$ & $8.26_{-1.70}^{+1.48}$ & $1.27_{-0.08}^{+0.09}$\\
  6603624 & $5512_{-46}^{+47}$ & $4.319_{-0.006}^{+0.006}$ & $0.32_{-0.10}^{+0.10}$ & $1.00_{-0.02}^{+0.02}$ & $1.15_{-0.01}^{+0.01}$ & $9.93_{-1.16}^{+1.16}$ & $1.09_{-0.04}^{+0.05}$\\
  8394589 & $5976_{-33}^{+74}$ & $4.306_{-0.009}^{+0.010}$ & $-0.32_{-0.10}^{+0.11}$ & $0.94_{-0.02}^{+0.04}$ & $1.14_{-0.01}^{+0.01}$ & $8.53_{-1.89}^{+1.13}$ & $1.48_{-0.06}^{+0.10}$\\
  9116461 & $6217_{-58}^{+56}$ & $4.314_{-0.011}^{+0.011}$ & $0.01_{-0.1}^{+0.10}$ & $1.16_{-0.04}^{+0.02}$ & $1.24_{-0.02}^{+0.02}$ & $3.13_{-0.73}^{+0.87}$ & $2.07_{-0.10}^{+0.12}$\\
  10079226 & $5869_{-80}^{+50}$ & $4.369_{-0.008}^{+0.008}$ & $0.21_{-0.090}^{+0.01}$ & $1.08_{-0.02}^{+0.04}$ & $1.13_{-0.01}^{+0.01}$ & $3.99_{-0.99}^{+1.17}$ & $1.37_{-0.08}^{0.07}$\\
  11133306 & $5894_{-70}^{+80}$ & $4.315_{-0.009}^{+0.010}$ & $-0.01_{-0.01}^{+0.01}$ & $1.04_{-0.04}^{+0.02}$ & $1.18_{-0.02}^{+0.02}$ & $6.20_{-1.33}^{+2.02}$ & $1.52_{-0.10}^{+0.09}$\\
  12068975 & $5929_{-33}^{+69}$ & $4.298_{-0.010}^{+0.014}$ & $-0.32_{-0.01}^{+0.10}$ & $0.92_{-0.02}^{+0.04}$ & $1.14_{-0.02}^{+0.02}$ & $9.45_{-1.51}^{+1.22}$ & $1.43_{-0.07}^{0.09}$\\
  \hline
  \it{Giant stars}&&&&&&&\\			
  \hline
   2010607 & $6117_{-64}^{+64}$ & $3.819_{-0.017}^{+0.018}$ & $0.10_{-0.10}^{+0.10}$ & $1.42_{-0.04}^{+0.04}$ & $2.44_{-0.07}^{+0.08}$ & $2.99_{-0.26}^{+0.21}$ & $7.58_{-0.66}^{0.49}$\\
   3344897 & $6234_{-63}^{+69}$ & $3.870_{-0.008}^{+0.009}$ & $0.010_{-0.010}^{+0.10}$ & $1.36_{-0.04}^{+0.04}$ & $2.25_{-0.03}^{+0.03}$ & $3.26_{-0.23}^{+0.23}$ & $6.80_{-0.37}^{0.45}$\\
     3438633 & $6076_{-57}^{+62}$ & $3.950_{-0.013}^{+0.013}$ & $-0.20_{-0.01}^{+0.10}$ & $1.14_{-0.02}^{+0.04}$ & $1.88_{-0.04}^{+0.04}$ & $5.41_{-0.43}^{+0.47}$ & $4.35_{-0.23}^{+0.26}$\\
     3456181 & $6234_{-59}^{+61}$ & $3.926_{-0.007}^{+0.008}$ & $0.01_{-0.01}^{+0.10}$ & $1.30_{-0.02}^{+0.04}$ & $2.07_{-0.02}^{+0.02}$ & $3.72_{-0.25}^{+0.29}$ & $5.82_{-0.3}^{+0.28}$\\
     3942719 & $5665_{-61}^{+53}$ & $3.833_{-0.019}^{+0.017}$ & $-0.25_{-0.09}^{+0.02}$ & $1.08_{-0.04}^{+0.04}$ & $2.09_{-0.07}^{+0.07}$ & $6.38_{-0.43}^{+0.86}$ & $4.04_{-0.32}^{0.34}$\\
     4049576 & $5810_{-60}^{+55}$ & $3.903_{-0.015}^{+0.015}$ & $ -0.17_{-0.09}^{+0.03}$ & $1.10_{-0.04}^{+0.02}$ & $1.94_{-0.04}^{+0.05}$ & $6.65_{-0.42}^{+0.51}$ & $3.89_{-0.24}^{+0.29}$\\
    4143755 & $5683_{-58}^{+56}$ & $4.095_{-0.008}^{+0.008}$ & $-0.40_{-0.10}^{+0.10}$ & $0.84_{-0.02}^{+0.02}$ & $1.37_{-0.02}^{+0.02}$ & $14.11_{-1.18}^{+1.30}$ & $1.77_{-0.10}^{0.09}$\\
    4165030 & $5678_{-57}^{+56}$ & $3.980_{-0.010}^{+0.011}$ & $-0.27_{-0.09}^{+0.09}$ & $0.94_{-0.02}^{+0.02}$ & $1.65_{-0.03}^{+0.03}$ & $10.40_{-0.81}^{+0.90}$ & $2.54_{-0.14}^{0.16}$\\
    4577484 & $5588_{-57}^{+54}$ & $3.861_{-0.013}^{+0.013}$ & $0.24_{-0.1}^{+0.05}$ & $1.26_{-0.02}^{+0.02}$ & $2.19_{-0.05}^{+0.05}$ & $5.10_{-0.32}^{+0.38}$ & $4.21_{-0.25}^{0.25}$\\
    4672403 & $5764_{-52}^{+59}$ & $3.994_{-0.009}^{+0.009}$ & $0.20_{-0.06}^{+0.09}$ & $1.18_{-0.04}^{+0.02}$ & $1.81_{-0.03}^{+0.02}$ & $6.44_{-0.40}^{+0.51}$ & $3.24_{-0.16}^{0.18}$\\
    4841753 & $5993_{-61}^{+49}$ & $3.906_{-0.017}^{+0.016}$ & $0.20_{-0.10}^{+0.10}$ & $1.30_{-0.02}^{+0.04}$ & $2.12_{-0.05}^{+0.05}$ & $4.14_{-0.24}^{+0.29}$ & $5.23_{-0.33}^{0.30}$\\
    5095159 & $5298_{-33}^{+71}$ & $3.711_{-0.008}^{+0.007}$ & $0.100_{-0.10}^{+0.10}$ & $1.48_{-0.06}^{+0.04}$ & $2.82_{-0.04}^{+0.05}$ & $2.71_{-0.25}^{+0.27}$ & $5.65_{-0.31}^{0.33}$\\
    5353186 & $6038_{-62}^{+62}$ & $3.885_{-0.010}^{+0.010}$ & $0.10_{-0.10}^{+0.10}$ & $1.32_{-0.02}^{+0.04}$ & $2.19_{-0.04}^{+0.03}$ & $3.89_{-0.26}^{+0.23}$ & $5.73_{-0.23}^{0.35}$\\
     5523099 & $5507_{-51}^{+57}$ & $3.794_{-0.010}^{+0.010}$ & $0.050_{-0.08}^{+0.04}$ & $1.26_{-0.04}^{+0.04}$ & $2.37_{-0.05}^{+0.04}$ & $4.57_{-0.27}^{+0.29}$ & $4.63_{-0.26}^{0.26}$\\
    5561278 & $5973_{-55}^{+63}$ & $3.960_{-0.012}^{+0.011}$ & $-0.06_{-0.04}^{+0.07}$ & $1.16_{-0.04}^{+0.02}$ & $1.87_{-0.03}^{+0.03}$ & $5.59_{-0.34}^{+0.68}$ & $4.01_{-0.18}^{0.22}$\\
    6064910 & $6173_{-60}^{+66}$ & $3.818_{-0.009}^{+0.009}$ & $-0.24_{-0.07}^{+0.04}$ & $1.24_{-0.04}^{+0.02}$ & $2.27_{-0.04}^{+0.03}$ & $4.03_{-0.42}^{+0.26}$ & $6.78_{-0.39}^{0.41}$\\
      6308642 & $5653_{-47}^{+61}$ & $3.797_{-0.007}^{+0.010}$ & $-0.13_{-0.07}^{+0.03}$ & $1.18_{-0.02}^{+0.02}$ & $2.28_{-0.03}^{+0.03}$ & $5.03_{-0.27}^{+0.31}$ & $4.79_{-0.20}^{0.21}$\\
     6520835 & $6030_{-59}^{+61}$ & $3.895_{-0.008}^{+0.007}$ & $0.00_{-0.1}^{+0.10}$ & $1.24_{-0.02}^{+0.02}$ & $2.09_{-0.03}^{+0.02}$ & $4.63_{-0.30}^{+0.32}$ & $5.11_{-0.28}^{0.29}$\\
     6587236 & $5916_{-60}^{+58}$ & $3.641_{-0.017}^{+0.018}$ & $-0.29_{-0.1}^{+0.09}$ & $1.42_{-0.06}^{+0.04}$ & $2.98_{-0.10}^{+0.08}$ & $2.43_{-0.21}^{+0.36}$ & $9.83_{-0.78}^{0.71}$\\
     6592305 & $6005_{-65}^{+59}$ & $3.870_{-0.006}^{+0.006}$ & $0.10_{-0.1}^{+0.1}$ & $1.32_{-0.04}^{+0.04}$ & $2.22_{-0.03}^{+0.02}$ & $3.92_{-0.25}^{+0.32}$ & $5.83_{-0.30}^{0.31}$\\
     6688822 & $5559_{-53}^{+56}$ & $3.860_{-0.009}^{+0.009}$ & $0.28_{-0.1}^{+0.10}$ & $1.28_{-0.02}^{+0.02}$ & $2.21_{-0.04}^{+0.03}$ & $5.06_{-0.29}^{+0.30}$ & $4.18_{-0.21}^{0.22}$\\
     6693861 & $5756_{-61}^{+52}$ & $3.835_{-0.010}^{+0.010}$ & $-0.25_{-0.09}^{+0.02}$ & $1.10_{-0.02}^{+0.02}$ & $2.10_{-0.04}^{+0.03}$ & $6.06_{-0.36}^{+0.32}$ & $4.37_{-0.26}^{0.23}$\\
      6766513 & $6169_{-69}^{+56}$ & $3.910_{-0.011}^{+0.011}$ & $-0.10_{-0.10}^{+0.10}$ & $1.24_{-0.04}^{+0.04}$ & $2.06_{-0.04}^{+0.04}$ & $4.24_{-0.42}^{+0.36}$ & $5.51_{-0.40}^{0.35}$\\
     6863041 & $5623_{-56}^{+53}$ & $3.814_{-0.010}^{+0.011}$ & $0.2_{-0.1}^{+0.08}$ & $1.34_{-0.02}^{+0.04}$ & $2.38_{-0.04}^{+0.05}$ & $4.04_{-0.24}^{+0.26}$ & $5.11_{-0.27}^{0.25}$\\
     7038145 & $5907_{-73}^{+60}$ & $3.820_{-0.008}^{+0.006}$ & $0.01_{-0.11}^{+0.10}$ & $1.32_{-0.04}^{+0.02}$ & $2.34_{-0.03}^{+0.03}$ & $3.79_{-0.06}^{+0.42}$ & $6.03_{-0.38}^{0.29}$\\
     7107778 & $5136_{-48}^{+61}$ & $3.649_{-0.009}^{+0.009}$ & $0.1_{-0.100}^{+0.10}$ & $1.50_{-0.10}^{+0.08}$ & $3.04_{-0.09}^{+0.08}$ & $2.62_{-0.42}^{+0.73}$ & $5.79_{-0.50}^{0.48}$\\
     7199397 & $5922_{-61}^{+56}$ & $3.757_{-0.007}^{+0.008}$ & $-0.1_{-0.10}^{+0.10}$ & $1.32_{-0.02}^{+0.06}$ & $2.54_{-0.04}^{+0.03}$ & $3.34_{-0.14}^{+0.17}$ & $7.08_{-0.29}^{0.36}$\\
     7264595 & $5671_{-47}^{+53}$ & $3.698_{-0.012}^{+0.011}$ & $-0.200_{-0.100}^{+0.09}$ & $1.34_{-0.04}^{+0.04}$ & $2.73_{-0.06}^{+0.06}$ & $3.15_{-0.27}^{+0.20}$ & $6.83_{-0.31}^{0.42}$\\
     7282890 & $6283_{-73}^{+68}$ & $3.874_{-0.008}^{+0.008}$ & $0.2_{-0.1}^{+0.10}$ & $1.48_{-0.03}^{+0.03}$ & $2.33_{-0.03}^{+0.03}$ & $2.69_{-0.25}^{+0.22}$ & $7.57_{-0.44}^{0.51}$\\
     7386523 & $5845_{-61}^{+56}$ & $3.883_{-0.018}^{+0.018}$ & $0.1_{-0.1}^{+0.1}$ & $1.26_{-0.04}^{+0.04}$ & $2.13_{-0.06}^{+0.07}$ & $4.74_{-0.43}^{+0.35}$ & $4.69_{-0.29}^{0.52}$\\
    7880676 & $6035_{-64}^{+55}$ & $3.944_{-0.009}^{+0.009}$ & $0.14_{-0.12}^{+0.06}$ & $1.26_{-0.04}^{+0.04}$ & $1.99_{-0.03}^{+0.03}$ & $4.50_{-0.40}^{+0.42}$ & $4.76_{-0.32}^{0.25}$\\
    8016496 & $5996_{-57}^{+60}$ & $3.952_{-0.009}^{+0.010}$ & $-0.06_{-0.04}^{+0.07}$ & $1.18_{-0.04}^{+0.02}$ & $1.90_{-0.03}^{+0.03}$ & $5.46_{-0.38}^{+0.47}$ & $4.18_{-0.21}^{0.22}$\\
    8019508 & $5995_{-57}^{+63}$ & $3.767_{-0.018}^{+0.017}$ & $0.09_{-0.10}^{+0.10}$ & $1.46_{-0.04}^{+0.04}$ & $2.61_{-0.07}^{+0.09}$ & $2.78_{-0.18}^{+0.26}$ & $7.99_{-0.55}^{0.55}$\\
     8045442 & $5902_{-58}^{+62}$ & $3.669_{-0.008}^{+0.010}$ & $0.09_{-0.1}^{+0.1}$ & $1.62_{-0.04}^{+0.02}$ & $3.08_{-0.04}^{+0.05}$ & $1.99_{-0.12}^{+0.10}$ & $10.32_{-0.51}^{0.57}$\\
     8493735 & $5887_{-60}^{+58}$ & $3.737_{-0.01}^{+0.014}$ & $-0.1_{-0.1}^{+0.1}$ & $1.36_{-0.04}^{+0.04}$ & $2.61_{-0.05}^{+0.06}$ & $3.18_{-0.26}^{+0.17}$ & $7.37_{-0.37}^{0.42}$\\
     8621637 & $5676_{-54}^{+55}$ & $3.961_{-0.010}^{+0.011}$ & $0.16_{-0.06}^{+0.06}$ & $1.16_{-0.04}^{+0.02}$ & $1.87_{-0.03}^{+0.03}$ & $6.80_{-0.55}^{+0.67}$ & $3.24_{-0.18}^{0.18}$\\
     8684730 & $5923_{-64}^{+59}$ & $3.927_{-0.013}^{+0.012}$ & $0.10_{-0.10}^{+0.10}$ & $1.24_{-0.04}^{+0.04}$ & $2.02_{-0.04}^{+0.04}$ & $4.94_{-0.48}^{+0.38}$ & $4.44_{-0.27}^{+0.40}$\\
     8802782 & $5868_{-60}^{+53}$ & $3.848_{-0.015}^{+0.015}$ & $0.2_{-0.1}^{+0.08}$ & $1.34_{-0.02}^{+0.04}$ & $2.30_{-0.05}^{+0.06}$ & $3.93_{-0.35}^{+0.13}$ & $5.58_{-0.26}^{0.47}$\\
     8817551 & $5762_{-54}^{+57}$ & $3.818_{-0.009}^{+0.008}$ & $0.2_{-0.1}^{+0.08}$ & $1.36_{-0.02}^{+0.04}$ & $2.39_{-0.04}^{+0.04}$ & $3.79_{-0.21}^{+0.23}$ & $5.70_{-0.32}^{0.29}$\\
     8868481 & $5589_{-56}^{+56}$ & $3.777_{-0.013}^{+0.013}$ & $0.01_{-0.01}^{+0.09}$ & $1.30_{-0.02}^{+0.04}$ & $2.45_{-0.06}^{+0.06}$ & $4.04_{-0.40}^{+0.27}$ & $5.31_{-0.30}^{0.32}$\\
    9335972 & $5715_{-53}^{+63}$ & $3.851_{-0.010}^{+0.009}$ & $0.15_{-0.10}^{+0.05}$ & $1.26_{-0.04}^{+0.04}$ & $2.22_{-0.04}^{+0.04}$ & $4.78_{-0.29}^{+0.30}$ & $4.74_{-0.24}^{0.24}$\\
    9592705 & $6068_{-58}^{+52}$ & $3.943_{-0.004}^{+0.003}$ & $0.20_{-0.10}^{+0.10}$ & $1.30_{-0.02}^{+0.02}$ & $2.02_{-0.02}^{+0.01}$ & $4.22_{-0.43}^{+0.32}$ & $5.00_{-0.24}^{+0.23}$\\
     9664694 & $6158_{-63}^{+57}$ & $3.808_{-0.011}^{+0.011}$ & $0.00_{-0.10}^{+0.10}$ & $1.38_{-0.04}^{+0.02}$ & $2.43_{-0.04}^{+0.04}$ & $3.14_{-0.18}^{+0.20}$ & $7.57_{-0.40}^{0.42}$\\
     9696358 & $6000_{-64}^{+55}$ & $3.915_{-0.030}^{+0.030}$ & $0.14_{-0.11}^{+0.06}$ & $1.28_{-0.04}^{+0.04}$ & $2.07_{-0.09}^{+0.09}$ & $4.36_{-0.39}^{+0.57}$ & $4.96_{-0.43}^{+0.55}$\\
     9715099 & $6139_{-67}^{+54}$ & $3.799_{-0.009}^{+0.009}$ & $0.10_{-0.1}^{+0.1}$ & $1.46_{-0.04}^{+0.04}$ & $2.53_{-0.04}^{+0.04}$ & $2.74_{-0.15}^{+0.24}$ & $8.08_{-0.40}^{0.43}$\\
    9757640 & $5516_{-54}^{+50}$ & $4.000_{-0.008}^{+0.009}$ & $0.29_{-0.06}^{+0.08}$ & $1.12_{-0.04}^{+0.02}$ & $1.75_{-0.02}^{+0.02}$ & $8.37_{-0.95}^{+0.71}$ & $2.53_{-0.08}^{0.14}$\\
   9778067 & $5906_{-60}^{+58}$ & $3.884_{-0.012}^{+0.014}$ & $-0.36_{-0.02}^{+0.01}$ & $1.06_{-0.02}^{+0.02}$ & $1.96_{-0.04}^{+0.04}$ & $6.30_{-0.13}^{+0.44}$ & $4.27_{-0.26}^{0.05}$\\
   9787965 & $6079_{-69}^{+56}$ & $3.929_{-0.016}^{+0.016}$ & $0.01_{-0.01}^{+0.10}$ & $1.26_{-0.04}^{+0.02}$ & $2.02_{-0.05}^{+0.05}$ & $4.50_{-0.46}^{+0.41}$ & $4.99_{-0.37}^{0.37}$\\
   9791157 & $5841_{-59}^{+58}$ & $3.940_{-0.009}^{+0.009}$ & $0.10_{-0.05}^{+0.10}$ & $1.20_{-0.04}^{+0.02}$ & $1.95_{-0.03}^{+0.03}$ & $5.68_{-0.40}^{+0.40}$ & $3.93_{-0.22}^{0.22}$\\
   10417911 & $5552_{-53}^{+56}$ & $3.955_{-0.017}^{+0.017}$ & $0.25_{-0.02}^{+0.04}$ & $1.16_{-0.02}^{+0.02}$ & $1.89_{-0.05}^{+0.05}$ & $7.02_{-0.47}^{+0.66}$ & $3.02_{-0.18}^{0.19}$\\
   10727922 & $5999_{-59}^{+58}$ & $3.946_{-0.019}^{+0.018}$ & $-0.11_{-0.10}^{+0.02}$ & $1.14_{-0.04}^{+0.04}$ & $1.89_{-0.05}^{+0.05}$ & $5.76_{-0.57}^{+0.54}$ & $4.14_{-0.24}^{+0.26}$\\
     10731424 & $6162_{-65}^{+62}$ & $3.773_{-0.019}^{+0.017}$ & $0.01_{-0.01}^{+0.09}$ & $1.44_{-0.04}^{+0.04}$ & $2.59_{-0.08}^{+0.08}$ & $2.70_{-0.24}^{+0.17}$ & $8.74_{-0.71}^{0.58}$\\
     10923629 & $6017_{-60}^{+62}$ & $3.819_{-0.008}^{+0.010}$ & $0.2_{-0.1}^{+0.08}$ & $1.42_{-0.02}^{+0.04}$ & $2.45_{-0.04}^{+0.03}$ & $3.05_{-0.18}^{+0.19}$ & $7.07_{-0.38}^{0.38}$\\
     11083308 & $6032_{-68}^{+59}$ & $3.911_{-0.011}^{+0.011}$ & $0.01_{-0.01}^{+0.09}$ & $1.24_{-0.02}^{+0.04}$ & $2.06_{-0.04}^{+0.04}$ & $4.60_{-0.49}^{+0.34}$ & $5.04_{-0.30}^{+0.35}$\\
     11138101 & $6148_{-62}^{+62}$ & $3.835_{-0.016}^{+0.014}$ & $0.01_{-0.01}^{+0.09}$ & $1.36_{-0.04}^{+0.04}$ & $2.34_{-0.06}^{+0.06}$ & $3.31_{-0.23}^{+0.21}$ & $7.08_{-0.52}^{0.49}$\\
     11193681 & $5594_{-58}^{+66}$ & $3.809_{-0.007}^{+0.008}$ & $0.20_{-0.01}^{+0.08}$ & $1.36_{-0.04}^{+0.02}$ & $2.40_{-0.03}^{+0.04}$ & $4.01_{-0.33}^{+0.24}$ & $5.10_{-0.26}^{0.24}$\\
   11771760 & $5862_{-62}^{+69}$ & $3.674_{-0.005}^{+0.010}$ & $0.01_{-0.01}^{+0.09}$ & $1.54_{-0.04}^{+0.04}$ & $2.99_{-0.06}^{+0.05}$ & $2.17_{-0.11}^{+0.20}$ & $9.53_{-0.50}^{0.47}$\\
     11817562 & $5974_{-59}^{+60}$ & $3.803_{-0.009}^{+0.009}$ & $0.2_{-0.01}^{+0.08}$ & $1.44_{-0.02}^{+0.02}$ & $2.50_{-0.04}^{+0.03}$ & $3.04_{-0.16}^{+0.18}$ & $7.11_{-0.35}^{0.38}$\\
     11919192 & $6368_{-69}^{+61}$ & $3.890_{-0.009}^{+0.009}$ & $-0.2_{-0.1}^{+0.09}$ & $1.28_{-0.02}^{+0.04}$ & $2.14_{-0.03}^{+0.03}$ & $3.48_{-0.32}^{+0.25}$ & $6.78_{-0.43}^{0.36}$\\
     12069127 & $6247_{-75}^{+46}$ & $3.891_{-0.010}^{+0.009}$ & $0.1_{-0.1}^{+0.1}$ & $1.42_{-0.06}^{+0.02}$ & $2.23_{-0.04}^{+0.03}$ & $3.02_{-0.16}^{+0.39}$ & $6.83_{-0.44}^{0.28}$\\

    \end{longtable}%
    \end{center}
\label{lastpage}


\clearpage

\begin{thebibliography}{99}

\bibitem[Appourchaux et al. 2012]{Appourchaux} Appourchaux, T., Chaplin, W. J., Garc\'{\i}a, R. A., et al. 2012, \aap, 543, a54
\bibitem[Basu et al. 2010]{Basu10} Basu, S., Chaplin, W. J., Elsworth, Y., 2010, \apj, 710, 1596
\bibitem[Basu et al. 2011]{Basu11} Basu, S., et al. 2011, \apj, 729, L10
\bibitem[Bahacall et al. 1995]{Bahacall95} Bahcall, J. N., Pinsonneault, M. H., \& Wasserburg, G. J., 1995, Reviews of
Modern Physics, 67, 781
\bibitem[Bedding \& Kjeldsen 2003]{Bedding03} Bedding, T. R., Kjeldsen, H., 2003, PASA, 20, 203
\bibitem[Bedding et al. 2011]{Bedding11} Bedding, T. R., et al. 2011, Nature, 471, 608
\bibitem[Bi et al. 2008]{Bi08} Bi, S. L., Basu, S., Li, L. H. 2008, \apj, 673, 1093
\bibitem[B\"{o}hm-Vitense 1958]{Bohm58} B\"{o}hm-Vitense, E. 1958, \zap, 46, 108
\bibitem[Christensen-Dalsgaard et al. 1993]{Christensen-Dalsgaard93} Christensen-Dalsgaard, J. 1993, ASP Conf. Ser. 42, Proc. GONG 1992
\bibitem[Chaplin et al. 2008]{Chaplin08} Chaplin, W. J., Houdek, G., Appourchaux, T., Elsworth, Y., New, R., Toutain, T., 2008, \aap, 485,813c
\bibitem[Chaplin et al. 2011]{Chaplin11} Chaplin, W. J., Kjeldsen, H., Christensen-Dalsgaard, J., et al. 2011, Science, 213, 6
\bibitem[Chaplin et al. 2014]{Chaplin14} Chaplin, W. J., Basu, S., Huber, D., et al. 2014, \apjs, 210, 1
\bibitem[Cui et al. 2012]{Cui12} Cui X. Q., Zhao Y. H., Chu Y. Q., et al. 2012, \raa, 12, 1197
\bibitem[De Cat et al. 2015]{De Cat Peter15} Peter, D. C., Fu, Jianning, Ren, Anbing, et al. 2015, \apjl, 220, 19
\bibitem[Deng et al. 2012]{Deng12} Deng L. C. et al., 2012, \raa, 12, 735
\bibitem[Demarque et al. 2004]{Demarque04} Demarque, P., Woo, J. H., Kim, Y. C., et al. 2004, \apjs, 155, 667
\bibitem[Demarque et al. 2008]{Demarque08} Demarque, P., Guenther, D. B., Li, L. H., Mazumdar, A., \& Straka, C. W. 2008, \apss, 316, 31
\bibitem[Dotter et al. 2008]{Dotter08} Dotter, A., et al. 2008, \apjs, 178, 89

\bibitem[Ferguson et al. 2005]{Ferguson05} Ferguson, J. W., Alexander, D. R., Allard, F., et al. 2005, \apj, 623, 585
\bibitem[1998]{Grevesse98} Grevesse, N., \& Sauval, A. J. 1998, \ssr, 85, 161
\bibitem[Gilliand et al. 2010]{G10} Gilliand, R. L., Jenkins, J. M., Borucki, W. J., et al. 2010, \apjl, 713, L160
\bibitem[Goudfrooij et al. 2009]{Goudfrooij09} Goudfrooij, P., Puzia, T. H., Kozhurina-Platais, V., et al. 2009, aj, 137, 4988
\bibitem[2015]{15} Huang, Y., Liu, X. W., Yuan, H. B., et al. 2015, \mnras, 454, 2863
\bibitem[Huber et al. 2014]{Huber14} Huber, D., Victor Silva Aguirre, Jaymie M. Matthews, et al. 2014, \apjs, 211, 2
\bibitem[Huber et al. 2013]{Huber13} Huber, D., Chaplin, W. J., Christensen-Dalsgaard, J., et al. 2013, \apj, 767, 127
\bibitem[Hekker et al. 2011]{Hekker11} Hekker, S., Y. Elsworth, J. De Ridder, et al. 2011, \aap, 525, a131
\bibitem[Hekker et al. 2013]{Hekker13} Hekker, S., Y. Elsworth, B. Mosser, et al. 2013, \aap, 556, a59
\bibitem[Iglesias et al. 1996]{Iglesias96} Iglesias, C. A., \& Rogers, F. J., 1996, \apj, 464, 943
\bibitem[Kjeldsen et al. 1995]{Kjeldsen95} Kjeldsen, H., \& Bedding, T. R. 1995, \aap, 293, 87
\bibitem[Kallinger et al. 2010]{Kallinger10} Kallinger, T., Mosser, B., Hekker, S., et al. 2010, \aap, 522, A1
\bibitem[Liu et al. 2014]{Liu14} Liu, X. W., Yuan, H. B., Luo, Z. Y., et al. 2014, in IAU Symposium 298 eds. S. Feltzing, G. Zhao, N. A. Walton, \&P. Whitelock, 310
\bibitem[Luo et al. 2012]{Luo12} Luo, A. L., Zhang, H. T., Zhao, Y. H., et al. 2012, \raa, 12, 1243
\bibitem[Luo et al. 2015]{Luo15} Luo, A. L., Zhao, Y. H., Zhao, G., et al. 2015, \raa, 15, 1095
\bibitem[Mackey et al. 2008]{Mackey08} Mackey, A. D., Broby Nielsen, P., Ferguson, A. M. N., et al. 2008, \apjl, 681, L17
\bibitem[Majewski et al. 2010]{Majewski10} Majewski S. R., Wilson J. C., Hearty F., Schiavon R. R., SkrutskieM.
F., 2010, in IAU Symposium, Vol. 265, IAU Symposium,
Cunha K., Spite M., Barbuy B., eds., pp. 480¨C481
\bibitem[Mosser et al. 2012]{Mosser12} Mosser, B., Y. Elsworth, S. Hekker, et al. 2012, \aap, 537, a30
\bibitem[Pinsonneault et al. 1990]{Pin90} Pinsonneault, M. H., Kawaler, S. D. \& Demarque, P. 1990, \apjs, 74, 501
\bibitem[Pinsonneault et al. 1992]{Pin92} Pinsonneault, M. H., Deliyannis, C. P. \& Demarque, P. 1992, \apjs, 78, 179
\bibitem[Ren et al. 2016]{Ren16} Ren, J. J., Liu, X. W., Xiang, M. S., et al. 2016, \raa, 16c, 9R
\bibitem[Rogers et al. 2002]{Rogers02} Rogers, F. J., \& Nayfonov, A. 2002, \apj, 576, 1064
\bibitem[Steinmetz et al. 2006]{Steinmetz06} Steinmetz M. et al., 2006, AJ, 132, 1645
\bibitem[Stello et al. 2010]{Stello10} Stello, D., Basu, S., Bruntt, H., et al. 2010, \apj, 713, L182
\bibitem[Stello et al. 2013]{Stello13} Stello, D., Huber, D., Bedding, T. R., et al. 2013, \apjl, 765, L41
\bibitem[1994]{Thoul94} Thoul, A. A., Bahcall, J. N., \& Loeb, A. 1994, \apj, 421, 828
\bibitem[Tian et al. 2015]{Tian15} Tian, Z. J., Bi, S. L., Bedding, Timothy R., \& Yang, W. M. 2015, \aap, 580, 44

\bibitem[Wang et al. 2016]{Wang16} Wang, L., Wang, W., Wu, Y., et al. 2016, arXiv, 1604, 05496
\bibitem[Wu et al. 2014]{Wu14} Wu, Y., Du, B., Luo, A. L., et al. 2014, IAUS, 306, 340
\bibitem[Xiang et al. 2015a]{Xiang15a} Xiang, M., S., Liu, X., W., Yuan, H., B., et al. 2015a, \mnras, 448, 822
\bibitem[Xiang et al. 2015b]{Xiang15b} Xiang, M. S., Liu, X. W., Yuan, H. B., et al. 2015b, \mnras, 448, 90
\bibitem[Yanny et al. 2009]{Yanny09} Yanny, B., Newberg, H. J., Johnson, J. A., et al., 2009, \apj, 700, 1282
\bibitem[Yang \& Meng 2010]{Yang10} Yang, W. M., \& Meng, X., 2010, NewA., 15, 367
\bibitem[Yang et al. 2013]{Yang13} Yang, W. M., Bi, S. L., Meng, X., Liu, Z., et al., 2013, \apj, 776, 112
\bibitem[Zhao et al. 2012]{Zhao12} Zhao, G., Zhao, Y. H., Chu, Y. Q., et al. 2012, \raa, 12, 723





\end{thebibliography}
\end{document}